\documentclass{elsart}

\usepackage{indentfirst}

\usepackage[latin1]{inputenc}
\usepackage{graphicx}
\usepackage{subfigure}
\usepackage{psfrag}

\usepackage{amsmath, amssymb}

\usepackage{xspace} 

\usepackage{color}
\definecolor{oneblue}{rgb}{0,0.0,0.75}

\newtheorem{proposition}{Proposition}

\newtheorem{remark}{Remark}

\newcommand{\od}[2]{\frac{d#1}{d#2}}
\newcommand{\pd}[2]{\frac{\partial#1}{\partial#2}}
\newcommand{\set}[1]{\left\{ #1 \right\}}
\newcommand{\vol}{\mathop{\mathrm{vol}}}
\newcommand{\area}{\mathop{\mathrm{area}}}

\newcommand{\norm}[1]{\left|\left|#1\right|\right|}

\def\a{\vec{a}}
\def\g{\vec{g}}
\def\u{\vec{u}}
\def\v{\vec{v}}
\def\vv{\mathbf{v}}
\def\div{\nabla\cdot}
\def\I{\mathbb{I}}
\def\w{\mathbf{w}}
\def\F{\mathcal{F}}
\def\S{\mathcal{S}}

\def\n{\vec{n}}
\def\A{\mathbb{A}}
\def\B{\mathbb{B}}
\def\N{\mathcal{N}}

\def\x{\vec{x}}
\def\k{\vec{k}}
\def\R{\mathbb{R}}
\def\rit{\mathbb{R}}
\def\Z{\mathbb{Z}}
\def\T{\mathcal{T}}
\def\O{\mathcal{O}}
\def\grad{\nabla}

\def\sgn{\mbox{sgn}}

\begin{document}

\begin{frontmatter}

\title{A compressible two-fluid model for the finite volume simulation of violent aerated flows. Analytical properties and numerical results}

\author[Dias]{Fr\'{e}d\'{e}ric Dias\corauthref{cor}}
\address[Dias]{LRC MESO, ENS Cachan, CEA DAM, 61 Av. du President Wilson, F-94230 Cachan, FRANCE}
\corauth[cor]{Corresponding author.}
\ead{Frederic.Dias@cmla.ens-cachan.fr}

\author[Dutykh]{Denys Dutykh}
\address[Dutykh]{LRC MESO, ENS Cachan, CEA DAM, 61 Av. du President Wilson, F-94230 Cachan, FRANCE}
\ead{Denys.Dutykh@cmla.ens-cachan.fr}

\author[JMG]{Jean-Michel Ghidaglia}
\address[JMG]{LRC MESO, ENS Cachan, CEA DAM, 61 Av. du President Wilson, F-94230 Cachan, FRANCE}
\ead{jmg@cmla.ens-cachan.fr}

\begin{abstract}
In the study of ocean wave impact on structures, one often uses Froude scaling since the dominant force is gravity. However the presence of trapped or entrained air in the water can significantly modify wave impacts. When air is entrained in water in the form of small bubbles, the acoustic properties in the water change dramatically and for example the speed of sound in the mixture is much smaller than in pure water, and even smaller than in pure air. While some work has been done to study small-amplitude disturbances in such mixtures, little work has been done on large
disturbances in air-water mixtures. We propose a basic two-fluid model in which both fluids share the same velocities. It is shown that this model can successfully mimic water wave impacts on coastal structures. Even though this is a model without interface, waves can occur. Their dispersion relation is discussed and the formal limit of pure phases (interfacial waves) is considered. The governing equations are discretized by a second-order finite volume method. Numerical results are presented. It is shown that this basic model can be used to study violent aerated flows, especially by providing fast qualitative estimates.
\end{abstract}


\end{frontmatter}


\section{Introduction}

One of the challenges in Computational Fluid Dynamics (CFD) is to determine efforts exerted by waves on structures,
especially coastal structures. The flows associated with wave impact can be quite complicated. In particular, wave breaking can lead to flows that cannot be described by models like {\it e.g.} the free-surface Euler or Navier--Stokes equations.
In a free-surface model, the boundary between the gas (air) and the liquid (water) is a surface. The liquid flow is assumed to be incompressible, while the gas is represented by a medium, above the liquid, in which the pressure is constant (the atmospheric pressure in general). Such a description is known to be valid for calculating the propagation in the open sea of waves with moderate amplitude, which do not break. Clearly it is not satisfactory when waves either break or hit coastal structures like offshore platforms, jetties, piers, breakwaters, etc.

Our goal here is to investigate a relatively simple two-fluid model that can handle breaking waves. It belongs to the family of averaged models, in the sense that even though the two fluids under consideration are not miscible, there exists a length scale $\epsilon$ such that each averaging volume (of size $\epsilon^3$)
contains representative samples of each of the fluids. Once the averaging process is performed, it is assumed that the two fluids share, locally, the same pressure, temperature and velocity. Such models are called homogeneous models in the literature. They can be seen as limiting cases of more general two-fluid models where the fluids could have different temperatures and velocities \cite{Ishii1975}. Let us explain why it can be assumed here that both fluids share the same temperatures and velocities. There are relaxation mechanisms that indeed tend to locally equalize these two quantities. Concerning temperatures, these are diffusion processes and provided no phenomenon is about to produce very strong gradients of temperature between the two fluids like {\it e.g} a nuclear reaction in one of the two fluids, one can assume that the time scale on which diffusion acts is much smaller than the time scale on which the flow is averaged. Similarly, concerning the velocities, drag forces tend to locally equalize the two velocities. Hence for flows in which the mean convection velocity is moderate (a scale of time is built with the mean convection velocity and a typical length scale) we are in the case where this time scale is much smaller than the time scale on which velocities are equalized through drag forces. Hence, in the present model, the partial differential equations, which express conservation of mass ($1$ per fluid), balance of momentum and total energy, read as follows:
\begin{eqnarray}\label{eq:massphys}
(\alpha^+\rho^+)_t  + \div(\alpha^+\rho^+\u) &=& 0, \\ \label{eq:massphys2}
  (\alpha^-\rho^-)_t  + \div(\alpha^-\rho^-\u) &=& 0, \\ \label{eq:momentumphys}
  (\rho\u)_t + \div\left(\rho\u\otimes\u + p\I\right) &=& \rho\g, \\
  \left(\rho E\right)_t + \div\left(\rho H\u\right) &=& \rho\g\cdot\u, \label{eq:energyphys}
\end{eqnarray}
where the superscripts $\pm$ are used to denote liquid and gas respectively. Hence $\alpha^+$ and $\alpha^-$ denote the volume fraction of liquid and gas, respectively, and satisfy the condition $\alpha^+ + \alpha^-=1$. We denote by $\rho^\pm$, $\u$, $p$, $e$ respectively the density of each phase, the velocity, the pressure, the specific internal  energy, $\g$ is the acceleration due to gravity,  $\rho := \alpha^+\rho^+ + \alpha^-\rho^-$ is the total density, $E = e + \frac12|\u|^2$ is the specific total energy, $H := E + {p}/{\rho}$ is the specific total enthalpy. In order to close the system, we assume that the pressure $p$ is given as a function of three parameters, namely $\alpha\equiv\alpha^+-\alpha^-$, $\rho$ and $e$:
\begin{equation}\label{EOS123}
p=\mathcal{P}(\alpha, \rho, e)\,.
\end{equation}
We shall discuss in Section \ref{modelstudy} how such a function $\mathcal{P}$ is determined once the two equations of state $p=\mathcal{P}^\pm(\rho^\pm, e^\pm)$ are known. Equations (\ref{eq:massphys})--(\ref{EOS123}) form a closed system that we shall use in order to simulate aerated flows.

The main purpose of this paper is to promote a general
point of view, which may be useful for various applications in ocean, offshore, coastal and arctic engineering.
One can say that the late Howell Peregrine was the first to make use of this approach. The influence of the presence of air in wave impacts is a difficult topic. While it is usually thought that the
presence of air softens the impact pressures, recent results show that the cushioning
effect due to aeration via the increased compressibility of the air-water mixture is not necessarily a dominant
effect \cite{Bullock2007}. First of all, air may become trapped or entrained in the water in different ways, for example as
a single bubble trapped against a wall, or as a column or cloud of small bubbles. In addition, it is not clear
which quantity is the most appropriate to measure impacts. For example some researchers
pay more attention to the pressure impulse than to pressure peaks. The pressure impulse is defined as the
integral of pressure over the short duration of impact. A long time ago, Bagnold \cite{Bagnold1939} noticed that the maximum pressure and impact duration differed from one identical wave impact to the next, even in carefully controlled laboratory
experiments, while the pressure impulse appears to be more repeatable. For sure, the simple one-fluid models which are commonly
used for examining the peak impacts are no longer appropriate in the presence of air. There are few studies dealing
with two-fluid models. An exception is the work by Peregrine and his collaborators. Wood et al. \cite{Wood2000}
used the pressure impulse approach to model a trapped air pocket. Peregrine \& Thiais \cite{Peregrine1996} examined the effect
of entrained air on a particular kind of violent water wave impact by considering a filling flow. Bullock et al. \cite{Bullock2001}
found pressure reductions when comparing wave impact between fresh and salt water where, due to the different properties
of the bubbles in the two fluids, the aeration levels being much higher in salt water than in fresh water. 
Bredmose \cite{Bredmose2005}
recently performed numerical experiments on a two-fluid system which has similarities with the one we will use below.

The novelty of the present paper is not the finite volume method used below but
rather the modelling of two-fluid flows. Since the model described below involves neither the tracking nor the
capture of a free surface, its integration is much less costly from the
computational point of view. We have chosen to report here on the stiffest case. Should the
viscosity effects become important, they can be taken into
account via e.g. a fractional step method. In fact, when viscous effects are important, the flow is
easier to capture from the numerical point of view.

The paper is organized as follows. Section \ref{modelstudy} provides an analytical study of the model. In Section
\ref{sec:prop} we show that in the limit where the function $\alpha$ is identically equal to $-1$ or $1$, equations (\ref{eq:massphys})--(\ref{eq:energyphys}) converge to the equations for the classical free-surface flow problems. We also study the dispersion relation of the new two-fluid model. Section \ref{modelnum} deals with the numerical discretization of this model via a finite volume method. Section \ref{modelres} is devoted to the presentation of the results of various numerical simulations. Finally a conclusion ends the paper.

\section{Analytical study of the model}
\label{modelstudy}

\subsection{The extended equation of state}

It is shown in this section how to determine the function $\mathcal{P}(\alpha,\rho,e)$ in Eq. (\ref{EOS123}) once the two equations of state $p=\mathcal{P}^\pm(\rho^\pm, e^\pm)$ are known.
We call Eq. (\ref{EOS123}) an extended EOS, since $\mathcal{P}(-1, \rho, e)=\mathcal{P}^-(\rho, e)$ and $\mathcal{P}(1, \rho, e)=\mathcal{P}^+(\rho, e)$, where
\begin{equation}\label{EOS567}
p^\pm=\mathcal{P}^\pm(\rho^\pm, e^\pm)\,,\quad T^\pm=\mathcal{T}^\pm(\rho^\pm, e^\pm)\,,
\end{equation}
are the EOS of each fluid. We will use the following prototypical example in this paper. Assume that the
fluid denoted by the superscript $-$ is an ideal gas:
\begin{equation}\label{eq:light}
  p^- = (\gamma^- - 1) \rho^- e^-, \qquad e^- = C_V^- T^-,
\end{equation}
while the fluid denoted by the superscript $+$ obeys to the stiffened gas law (Tait's law) \cite{Godunov1979}:
\begin{equation}\label{eq:heavy}
  p^+ + \pi^+ = (\gamma^+ - 1) \rho^+ e^+, \qquad e^+ = C_V^+T^+ + \frac{\pi^+}{\gamma^+ \rho^+},
\end{equation}
where $\gamma^\pm$, $C_V^\pm$, and $\pi^+$ are constants. For example, pure water is well described in the vicinity of the normal conditions by taking $\gamma^+ = 7$ and $\pi^+ = 2.1\times10^9$ Pa.

Let us now return to the general case. In order to find the function $\mathcal{P}$, there are three given quantities: $\alpha\in[-1,1]$\,, $\rho>0$ and $e>0\,$. Then one solves for the four unknowns $\rho^\pm\,, e^\pm$ the following system of four nonlinear equations:
\begin{eqnarray}\label{nonlin1}
  (1+\alpha)\rho^++(1-\alpha)\rho^- &=& 2\rho\,, \\\label{nonlin2}
  (1+\alpha)\rho^+e^++(1-\alpha)\rho^-e^- &=& 2\rho\,e\,, \\\label{nonlin3}
  \mathcal{P}^+(\rho^+, e^+)-\mathcal{P}^-(\rho^-, e^-) &=& 0\,, \\\label{nonlin4}
 \mathcal{T}^+(\rho^+, e^+)-\mathcal{T}^-(\rho^-, e^-) &=& 0\,.
\end{eqnarray}
For given values of the pressure $p>0$ and the temperature $T>0$, we denote by $\mathcal{R}^\pm(p,T)$ and $\mathcal{E}^\pm(p,T)$ the solutions $(\rho^\pm, e^\pm)$ to:
\begin{equation}\label{inversion1}
\mathcal{P}^\pm(\rho^\pm, e^\pm)=p\,,\quad \mathcal{T}^\pm(\rho^\pm, e^\pm)=T\,,
\end{equation}
and then:
\begin{eqnarray}\label{inversion2}
\rho&=&\frac{1+\alpha}{2}\mathcal{R}^+(p,T)+\frac{1-\alpha}{2}\mathcal{R}^-(p,T)\,,\\
\label{inversion3}
\rho\,e&=&\frac{1+\alpha}{2}\mathcal{R}^+(p,T)\,\mathcal{E}^+(p,T)+\frac{1-\alpha}{2}\mathcal{R}^-(p,T)\,\mathcal{E}^-(p,T)\,.
\end{eqnarray}
Finally the inversion of this system of equations leads to $p=\mathcal{P}(\alpha,\rho,e)$ and $T=\mathcal{T}(\alpha,\rho,e)$.\\

Concerning the prototypical case, the following generalization of (\ref{eq:light}) is considered:
\begin{equation}\label{eq:light1}
  p^- + \pi^-= (\gamma^- - 1) \rho^- e^-, \qquad e^- = C_V^- T^-+ \frac{\pi^-}{\gamma^- \rho^-}\,.
\end{equation}
Introducing $\gamma$ and $\pi$ defined by
\begin{equation}\label{gamma}
\frac{2}{\gamma(\alpha)-1} =\frac{1+\alpha}{\gamma^+-1}+\frac{1-\alpha}{\gamma^--1}\,,
\end{equation}
\begin{equation}\label{pi}
\frac{2\,\pi(\alpha)}{\gamma(\alpha)-1} =\frac{1+\alpha}{\gamma^+-1}\pi^++\frac{1-\alpha}{\gamma^--1}\pi^-\,,
\end{equation}
Eq. (\ref{inversion2}) and (\ref{inversion3}) then lead to
\begin{eqnarray}\label{EOSpression}
\mathcal{P}(\alpha, \rho, e) & = & (\gamma(\alpha)-1)\rho\,e-\pi(\alpha)\,, \\
\label{EOStemperature}
\mathcal{T}(\alpha, \rho, e) & = & \frac{\rho\,e-(\lambda^+(\alpha)\pi^++\lambda^-(\alpha)\pi^-)}{\rho \, C_V(\alpha)}\,,
\end{eqnarray}
where
\begin{equation}\label{defCV}
\left(\frac{1+\alpha}{C_V^+(\gamma^+-1)}+\frac{1-\alpha}{C_V^-(\gamma^--1)}\right)C_V(\alpha)=
\frac{1+\alpha}{\gamma^+-1}+\frac{1-\alpha}{\gamma^--1}\,,
\end{equation}
\begin{equation}\label{deflambda}
\lambda^\pm(\alpha)\equiv\frac{1\pm\alpha}{2(\gamma^\pm-1)}
\left(1-\frac{C_V(\alpha)}{\gamma^\pm C_V^\pm}\right)\,.
\end{equation}
One can easily check that one recovers the equations of state for each fluid in the limits $\alpha \to \pm 1$.

\subsection{An hyperbolic system of conservation laws}

In this section, we assume that the system of equations is solved in $\rit^2$, having in mind the numerical
computations performed below. However the extension to 3D is trivial. 
The system (\ref{eq:massphys})--(\ref{eq:energyphys}) can be written as
\begin{equation}\label{syst_abst}
  \pd{\w}{t} + \div\F(\w) = \S(\w),
\end{equation}
where
\begin{equation}\label{eq:consvars}
  \w = (w_i)_{i=1}^{5} := (\alpha^+\rho^+, \alpha^-\rho^-,\;\;\rho u_1,\;\;\rho u_2,\;\;\rho E)\,,
\end{equation}
and, for every $\n\in\rit^2$,
\begin{equation}\label{flux1}
\F(\w)\cdot\n = (\alpha^+\rho^+\u\cdot \n, \alpha^-\rho^-\u\cdot \n, \rho \u\cdot \n u_1+ p n_1, \rho \u\cdot \n u_2 + pn_2, \rho H \u\cdot \n)\,,
\end{equation}
\begin{equation}\label{source}
\S(\w)=(0,0,\rho g_1,\rho g_2,\rho\g\cdot\u)\,.
\end{equation}
The Jacobian matrix $\A(\w)\cdot\n$ is defined by
\begin{equation}\label{mat_jacob}
\A(\w)\cdot\n=\pd{(\F(\w) \cdot\n)}{\w}\,.
\end{equation}

In order to compute $\A(\w)\cdot\n$, one writes Eq. (\ref{flux1}) for $\F(\w)\cdot\n$ in terms of $\w$ and $p$:
\begin{multline}\label{eq:normalAdvflux}
  \F(\w)\cdot\n =  \Bigl(w_1\frac{w_3n_1 + w_4n_2}{w_1+w_2}, w_2\frac{w_3n_1 + w_4n_2}{w_1+w_2}, w_3\frac{w_3n_1 + w_4n_2}{w_1+w_2} + p n_1, \\ w_4\frac{w_3n_1 + w_4n_2}{w_1+w_2} + pn_2, (w_5 + p)\frac{w_3n_1 + w_4n_2}{w_1+w_2}\Bigr)\,.
\end{multline}
The Jacobian matrix (\ref{mat_jacob}) then has the following expression:
\begin{eqnarray*}
 \A(\w)\cdot\n & = & \\
 & & \hspace{-3.5cm} \begin{pmatrix}
   u_n\frac{\alpha^-\rho^-}{\rho} & -u_n\frac{\alpha^+\rho^+}{\rho} & \frac{\alpha^+\rho^+}{\rho} n_1 & \frac{\alpha^+\rho^+}{\rho} n_2 & 0 \\
   -u_n\frac{\alpha^-\rho^-}{\rho} & u_n\frac{\alpha^+\rho^+}{\rho} & \frac{\alpha^-\rho^-}{\rho} n_1 & \frac{\alpha^-\rho^-}{\rho} n_2 & 0 \\
   -u_1 u_n+\pd{p}{w_1}n_1 & -u_1 u_n+\pd{p}{w_2}n_1 & u_n + u_1 n_1 + \pd{p}{w_3}n_1 & %
   u_1 n_2 + \pd{p}{w_4} n_1 & \pd{p}{w_5} n_1 \\
   -u_2 u_n+\pd{p}{w_1}n_2 & -u_2 u_n+\pd{p}{w_2}n_2 & u_2 n_1 + \pd{p}{w_3}n_2 & %
   u_n + u_2 n_2 + \pd{p}{w_4} n_2 & \pd{p}{w_5} n_2 \\
   u_n\bigl(\pd{p}{w_1} - H\bigr) & u_n\bigl(\pd{p}{w_2} - H\bigr) & %
   u_n\pd{p}{w_3} + H n_1 & u_n\pd{p}{w_4} + H n_2 & u_n\bigl(1 + \pd{p}{w_5}\bigr) \\
 \end{pmatrix}\,,
\end{eqnarray*}
where $u_n = \u \cdot \n.$

Let us now compute the five derivatives ${\partial p}/{\partial w_i}$. A systematic way of doing it is to introduce a set of five independent physical variables and here we shall take:
\begin{equation}\label{varphys}
\varphi_1=\alpha, \quad \varphi_2=p, \quad \varphi_3=T, \quad \varphi_4=u_1, \quad \varphi_5=u_2\,.
\end{equation}
The expressions of the $w_i's$ in terms of the $\varphi_j's$ are algebraic and explicit. Hence the Jacobian matrix $\partial{w_i}/\partial{\varphi_j}$ can be easily computed. Since $\partial{\varphi_j}/\partial{w_i}$ is its inverse matrix, one finds easily with the help of a computer algebra program that
\begin{eqnarray}
  \pd{p}{w_1} = \frac{\Gamma-1}{2}(u_1^2+u_2^2)+\alpha^-\rho^-\chi^-\,, \\
  \pd{p}{w_2} = \frac{\Gamma-1}{2}(u_1^2+u_2^2)+\alpha^+\rho^+\chi^+\,,
\end{eqnarray}
\begin{equation}\label{deriv_de_p}
 \pd{p}{w_3}=-(\Gamma-1)u_1\,,\quad \pd{p}{w_4}=-(\Gamma-1)u_2\,,\quad \pd{p}{w_5}=\Gamma-1\,,
\end{equation}
where
\begin{equation}\label{chipm}
\chi^\mp=\frac{1}{\rho^\pm}\frac{(c^\mp_s)^2}{\gamma^\mp-1}-\frac{1}{\rho^\mp}\frac{(c^\pm_s)^2}{\gamma^\pm-1}\,,\quad
\chi^++\chi^-=0\,,
\end{equation}
\begin{equation}\label{cscarre}
(c^\pm_s)^2\equiv C_V^\pm\gamma^\pm(\gamma^\pm-1)T=\frac{\gamma^\pm p+\pi^\pm}{\rho^\pm}\,,
\end{equation}
\begin{equation}\label{defGamma}
\Gamma-1\equiv(\gamma(\alpha)-1)\frac{\rho c_s^2}{\gamma(\alpha)p+\pi(\alpha)}\,.
\end{equation}
In Eq. (\ref{defGamma}), we have introduced the speed of sound of the mixture $c_s$, defined by
\begin{equation}
\frac{1}{\rho c_s^2}=\frac{(1+\alpha)\gamma^+}{2\rho^+(c^+_s)^2}+\frac{(1-\alpha)\gamma^-}{2\rho^-(c^-_s)^2}-
\frac{1}{\rho a^2}\,,
\label{def1surrhoc2}
\end{equation}
with
\begin{equation}
\rho a^2\equiv \frac{(1+\alpha)\rho^+(c_s^+)^2}{2(\gamma^+-1)}+\frac{(1-\alpha)\rho^-(c_s^-)^2}{2(\gamma^--1)}\,.
\label{def1sura2}
\end{equation}
Then one can show that the Jacobian matrix $ \A(\w)\cdot\n $ has three distinct eigenvalues:
\begin{equation}
  \lambda_1 = u_n - c_s, \quad
  \lambda_{2,3,4} = u_n, \quad
  \lambda_5 = u_n + c_s,
\end{equation}
and is diagonalizable on $\rit$. The expression of a set of eigenvectors can be obtained by using a computer algebra program.
\begin{remark}
If $\pi^+=0$ and $\pi^-=0$, then $c_s^2=\frac{\gamma(\alpha)p}{\rho}$ and $a^2=\frac{c_s^2}{\gamma(\alpha)-1}$.
\end{remark}
\begin{remark}
The left hand side of (\ref{def1surrhoc2}) is positive since $\rho a^2$ is bounded from below by $\frac{(1+\alpha)\rho^+(c^+_s)^2}{2\gamma^+}+\frac{(1-\alpha)\rho^-(c^-_s)^2}{2\gamma^-}$
\end{remark}
\subsection{Evolution equations for the physical variables}

The system of conservation laws (\ref{eq:massphys})--(\ref{eq:energyphys}) can be transformed into a set of evolution equations for the physical variables. Let us introduce the entropy function $s(\x,t)$ defined by (compare with Eq. (\ref{nonlin2}))
$$ 2\rho\,s = (1+\alpha)\rho^+s^++(1-\alpha)\rho^-s^-. $$
\begin{proposition}
Continuous solutions to (\ref{eq:massphys})--(\ref{eq:energyphys}) satisfy
\begin{eqnarray}\label{eq:vitesse}
\u_t + \u\cdot\nabla\u + \frac{1}{\rho}\nabla p &=& \g\,, \\\label{eq:pression}
  p_t + \u\cdot\nabla p + \rho c_s^2\div \u &=& 0\,, \\\label{eq:alpha}
  \alpha_t + \u\cdot\nabla \alpha + (1-\alpha^2)\,\delta\,\div \u &=& 0\,, \\
  s_t + \u\cdot\nabla s &=& 0\,, \label{eq:entropie}
  \end{eqnarray}
  where $c_s^2$ is given by (\ref{def1surrhoc2})-(\ref{def1sura2}) and $\delta$ is given by
\begin{equation}\label{def_delta}
\delta\equiv \frac{1}{2}\frac{\rho c_s^2(\gamma^-\pi^+-\gamma^+\pi^-)}{\rho^+\rho^-(c_s^+)^2(c_s^-)^2}\,.
\end{equation}
\end{proposition}

\begin{remark}
For pure fluids ($\alpha=\pm 1$), Eq. (\ref{eq:alpha}) is no longer relevant and $\delta$ is not needed. One can check
that the speed of sound $c_s$ is then equal to the expected speed of sound ($c_s^+$ or $c_s^-$) for pure fluids.
\end{remark}

The balance of entropy (\ref{eq:entropie}) comes from the balance
\begin{equation}\label{eq:entropy}
 (\rho s)_t  + \div (\rho s \u) = 0. 
\end{equation}
Adding together Eqs (\ref{eq:massphys}) and (\ref{eq:massphys2}) leads to
\begin{equation}\label{eq:total_mass}
\rho_t + \div (\rho \u) =  0.
\end{equation}
Combining Eqs (\ref{eq:entropy}) and (\ref{eq:total_mass}) leads to Eq. (\ref{eq:entropie}).

\begin{remark}
Subtracting Eq. (\ref{eq:massphys}) from (\ref{eq:massphys2}) leads to
\begin{equation}\label{chi}
(\rho\chi)_t + \div (\rho\chi \u) =  0\,, \quad \mbox{with} \;\; \chi = \frac{\alpha^+\rho^+-\alpha^-\rho^-}
{\rho}\,.
\end{equation}
In the case of smooth solutions, we obtain that 
$$ \chi_t + \u\cdot\nabla\chi = 0 \,, $$
which is an alternative to Eq. (\ref{eq:alpha}). 
\end{remark}

\section{Properties of the model}
\label{sec:prop}

\subsection{Basic state}

From now on, we denote the set of equations (\ref{eq:massphys})--(\ref{eq:energyphys}) by (E).
In order to study small perturbations around basic smooth and stationary solutions, it
is more convenient to use the general set of equations (E) rewritten
in the physical variables $\alpha$, $\u$, $p$ and $s$: see Eq. (\ref{eq:vitesse})--(\ref{eq:entropie}).

The two-fluid model (E) describes the evolution of mixtures. It can be used for example to study waves along a diffuse
interface between a gas and a liquid. In order to find the dispersion relation for such waves,
one first looks for rest states. There is an infinity of such rest states. Then, the governing equations are linearized around a special class of rest states. Here the situation is considerably complicated by the fact that the stationary solutions are not uniform in space. Thus, we come up with a linear system of partial differential equations which have non-constant coefficients.

The steady state will be denoted by $\underline{\alpha}^\pm$, $\underline{\rho}^\pm$, $\underline{p}$, $\underline{\u}$
and $\underline{s}$. The special class of solutions we are looking for are motionless, uniform in the horizontal coordinates and continuously stratified
in the vertical direction:
\begin{equation*}
  \underline{\alpha}^\pm = \underline{\alpha}^\pm (z), \quad
  \underline{\rho}^\pm = \underline{\rho}^\pm (z), \quad
  \underline{p} = \underline{p} (z), \quad
  \underline{\u} = \vec{0}, \quad \underline{s}=\underline{s}(z).
\end{equation*}
In this case, Eqs (\ref{eq:vitesse})--(\ref{eq:entropie}) become
\begin{equation} \label{hydrop}
\frac{d\underline{p}}{dz} = -\underline{\rho}(z)g\,.
\end{equation}
In the case where one makes the assumption that the mixture density is constant,
\begin{equation*}
  \underline{\rho} (z) = \frac{1}{2}(1+\underline{\alpha})\underline{\rho}^+ +
  \frac{1}{2}(1-\underline{\alpha})\underline{\rho}^- \equiv \rho_0,
\end{equation*}
where $\rho_0$ is some positive constant, the solution to (\ref{hydrop}) is
\begin{equation} \label{hydrop2}
  \underline{p}(z) = p_0 - \rho_0 g z,
\end{equation}
where $p_0$ is a constant.

Equation (\ref{hydrop2}) combined with (\ref{EOSpression}) leads to
\begin{equation} \label{pressuregeneral}
p_0 - \rho_0 g z = (\gamma(\underline{\alpha})-1)\rho_0\,\underline{e}-\pi(\underline{\alpha})\,,
\end{equation}
and there are infinitely many solutions. Since we have imposed $\underline{\rho}(z)=\rho_0$, it makes sense
to also impose a temperature which does not depend on $z$: $\underline{T}(z)=T_0$. Otherwise, there would be some
motion due to convection ($\u \neq 0$). Thus, Eq. (\ref{EOStemperature}) leads to the equation
\begin{equation}
\underline{e} = C_V(\underline{\alpha}) T_0 + \frac{\lambda^+(\underline{\alpha})\pi^++\lambda^-(\underline{\alpha})\pi^-}{\rho_0},
\end{equation}
which can be used to obtain an equation for $\underline{\alpha}$ when combined with Eq. (\ref{pressuregeneral}).


\subsection{Linearized two-fluid equations}

In this section we linearize the governing equations around the basic steady state derived in the previous section. We write down the following perturbation of the stationary solution:
$$ \alpha = \underline{\alpha}(z) + 2\beta + \ldots \,, \quad
  p = \underline{p}(z) + q + \ldots \,, \quad
  \u = \vec{0} + \v + \ldots \,, \quad s = \underline{s}(z) + \sigma + \ldots \,,
$$
where the vector $\v$ has the three components $(v_1, v_2, v_3)$.

From Eqs (\ref{eq:pression})--(\ref{eq:entropie}), it is straightforward to check that $q$, $\beta$ and $\sigma$
satisfy the equations
\begin{eqnarray}
  \pd{q}{t} + \v\cdot\grad\underline{p}(z) +
  {\rho}_0\,\underline{c}_s^2 \div\v &=& 0\,, \label{eq:pilinear} \\
    \pd{\beta}{t} + \v\cdot\grad\underline{\alpha}(z) +
  \frac{1}{2}(1-\underline{\alpha}^2)\underline{\delta}\div\v &=& 0\,, \label{eq:betalinear} \\
  \pd{\sigma}{t} + \v\cdot\grad\underline{s}(z) & = & 0\,. \label{eq:sigmalinear}
\end{eqnarray}

In order to obtain the velocity perturbation equation, one needs to compute carefully the density perturbation. Using that
$$ \frac{d\rho^\pm}{dz} = \frac{\gamma^\pm}{(c_s^\pm)^2} \frac{dp}{dz} \quad \mbox{(recall that} \;\; \frac{dT}{dz}=0) \,, $$
\begin{eqnarray*}
  \rho & = & \frac{1}{2}(1+\underline{\alpha}(z) + 2\beta + \ldots) \mathcal{R}^+(\underline{p}(z) + q + \ldots,T_0) \\
& & \hspace{1cm} + \frac{1}{2}
  (1-\underline{\alpha}(z) - 2\beta + \ldots) \mathcal{R}^-(\underline{p}(z) + q + \ldots,T_0) \\
  & = & \frac{1}{2}(1+\underline{\alpha}(z) + 2\beta + \ldots)
  \left(\underline{\rho}^+ + \frac{\gamma^+ q}{(c_s^+)^2} + \ldots \right) \\
& & \hspace{1cm} + \frac{1}{2}
  (1-\underline{\alpha}(z) - 2\beta + \ldots) \left(\underline{\rho}^- + \frac{\gamma^- q}{(c_s^-)^2} + \ldots\right) \\
& = & \rho_0 + \beta(\underline{\rho}^+ - \underline{\rho}^-) +
  \frac{1}{2}\left(\frac{(1+\underline{\alpha})\gamma^+}{(c_s^+)^2} +
  \frac{(1-\underline{\alpha})\gamma^-}{(c_s^-)^2}\right)q + \O(\norm{\pi}^2 + \norm{\beta}^2).
\end{eqnarray*}

Having computed $\rho$, it is easy to write the equation for $\v$ from Eq. (\ref{eq:vitesse}):
\begin{equation}\label{eq:vlinear}
  \pd{\v}{t} + \frac{1}{\rho_0} \grad q = \left[ \frac{\underline{\rho}^+ - \underline{\rho}^-}{\rho_0}\beta +
\frac{1}{2\rho_0}\left(\frac{(1+\underline{\alpha})\gamma^+}{(c_s^+)^2} +
  \frac{(1-\underline{\alpha})\gamma^-}{(c_s^-)^2}\right) q \right] \g \,.
\end{equation}


\subsection{Dispersion relation}

From now on, the following notation will be used:
\begin{eqnarray} \nonumber
  & & a(z) := {\rho}_0\,\underline{c}_s^2\,, \quad
  b(z) := \frac{1}{2}(1-\underline{\alpha}^2)\underline{\delta} \,, \\ \label{abcd}
  & & c(z) := \frac{\underline{\rho}^+ - \underline{\rho}^-}{\rho_0} \,, \quad
d(z) := \frac{1}{2\rho_0}\left(\frac{(1+\underline{\alpha})\gamma^+}{(c_s^+)^2}+\frac{(1-\underline{\alpha})\gamma^-}{(c_s^-)^2}\right) \,.
\end{eqnarray}

\subsubsection{Case without gravity}

Consider first the simple case where the acceleration due to gravity is absent: $\g=0$. It represents a major simplification since in this case we recover from (\ref{eq:pilinear})--(\ref{eq:vlinear}) a system of partial differential equations with constant coefficients:
\begin{eqnarray*}
  \pd{q}{t} + {\rho}_0\,\underline{c}_s^2 \div\v &=& 0\,, \\
  \pd{\beta}{t} + \frac{1}{2}(1-\underline{\alpha}^2)\underline{\delta}\div\v &=& 0\,, \\
  \pd{\sigma}{t} & = & 0\,, \\
  \pd{\v}{t} + \frac{1}{\rho_0} \grad q &=& 0.
\end{eqnarray*}
This system can be written in abstract form as
\begin{equation}\label{eq:conslaw}
  \pd{\w}{t} + \sum_{i=1}^3 \B_i\pd{\w}{x_i} = 0,
\end{equation}
where $\w = (q, \beta, \sigma, \v)$. We look for plane wave solutions:
$$
	\w(\x,t) = \w_0 e^{i(\k\cdot\x - \omega t)}, \quad \x = (x_1,x_2,x_3) \;\; \mbox{with} \;\; x_3=z, \quad \k = (k_1, k_2, k_3).
$$
Substituting this ansatz into equation (\ref{eq:conslaw}) yields
\begin{equation*}
  \B(\k)\w_0 = \omega\w_0\,, \quad \mbox{with} \;\; \B(\k) = \sum_{i=1}^3 k_i \B_i 
\end{equation*}
or
\begin{equation*}
 \B(\k) = \begin{pmatrix}
   0 & 0 & 0 & a_0 k_1 & a_0 k_2 & a_0 k_3 \\
   0 & 0 & 0 & b_0 k_1 & b_0 k_2 & b_0 k_3 \\
   0 & 0 & 0 & 0 & 0 & 0 \\
   \frac{k_1}{\rho_0} & 0 & 0 & 0 & 0 & 0 \\
   \frac{k_2}{\rho_0} & 0 & 0 & 0 & 0 & 0 \\
   \frac{k_3}{\rho_0} & 0 & 0 & 0 & 0 & 0 \\
 \end{pmatrix}, \quad a_0=\rho_0 \underline{c}_s^2, \;\; b_0=\frac{1}{2}(1-\underline{\alpha}^2)\underline{\delta}\,.
\end{equation*}
It means that $\omega$ is an eigenvalue and $\w_0$ the corresponding eigenvector of the matrix $\B_k$.

The dispersion relation of the system (\ref{eq:conslaw}) is given by its characteristic polynomial
\begin{equation*}
  \omega^4\left(\omega^2 - \frac{a_0}{\rho_0}|\k|^2\right) = 0 \quad \mbox{or} \quad
  \omega^4\left(\omega^2 - \underline{c}_s^2|\k|^2\right) = 0\,.
\end{equation*}
Note that here we have not considered any boundary conditions and that the vertical direction does not play
any particular role. This is why we have been looking for perturbations which are periodic in all three directions.
In fact there is no dispersion in the acoustic waves we have found.

\begin{remark}
The computations performed above are in full agreement with the computations (\ref{syst_abst})--(\ref{mat_jacob}). The matrix $\B$
is similar to $\A(0)$. Note that we obtained here 6 eigenvalues as opposed to 5 in Section 2.2. This is only due
to the fact that we performed the computations in 3D in this section as opposed to 2D in Section 2.2.
\end{remark}

\subsubsection{General case with gravity}

Let us now consider the general situation where $g$ is different from zero. We drop the equation for $\sigma$ since
(\ref{eq:pilinear})--(\ref{eq:sigmalinear}) is a strictly ``triangular'' linear system of PDEs. As above we look for periodic solutions of the form
\begin{equation}\label{eq:zdepending}
  \begin{pmatrix}
        q \\ \beta \\ \v
  \end{pmatrix} =
  \begin{pmatrix}
    \hat q \\
    \hat\beta \\
    \hat\vv
  \end{pmatrix} (z) e^{i(\k\cdot\x - \omega t)}.
\end{equation}
In this case $\x$ and $\k$ have only horizontal components:
\begin{equation*}
  \x = (x_1, x_2), \quad \k = (k_1, k_2),
\end{equation*}
and $\hat\vv = (\hat v_1, \hat v_2, \hat w)$. The main difference with the previous case is that the amplitude now depends on the vertical coordinate $z$. It makes the analysis more complicated.

Substituting the expression (\ref{eq:zdepending}) into Eqs (\ref{eq:pilinear})--(\ref{eq:vlinear}) except for
(\ref{eq:sigmalinear}) yields the
following  system of ordinary differential equations:
\begin{eqnarray*}
  -i\omega\hat q - \rho_0 g\hat w +
  a(z)\Bigl(i\k\cdot\hat v_{1,2} + \od{\hat w}{z}\Bigr) &=& 0, \\
    -i\omega\hat\beta + \od{\underline{\alpha}}{z}\hat w +
  b(z)\Bigl(i\k\cdot\hat v_{1,2} + \od{\hat w}{z}\Bigr) &=& 0, \\
  -i\omega\hat v_{1,2} + i\k\frac{\hat q}{\rho_0} &=& \vec{0}, \\
  -i\omega\hat w + \frac{1}{\rho_0}\od{\hat q}{z} + c(z)g\hat\beta +
  d(z)g\hat q &=& 0\,.
\end{eqnarray*}

The third equation yields the horizontal divergence of $\v$ in Fourier space in terms of the pressure perturbation $\hat q$:
\begin{equation*}
  i\k\cdot\hat v_{1,2} = i\frac{|\k|^2}{\omega}\frac{\hat q}{\rho_0}.
\end{equation*}

Another algebraic identity can be obtained if we multiply the first equation by $b(z)$, the second one by $a(z)$ and subtract them:
\begin{equation*}
  -i\omega\hat\beta a(z) + \Bigl(\od{\underline{\alpha}}{z}a(z) + \rho_0 g b(z)\Bigr)\hat w + i\omega\hat q b(z) = 0.
\end{equation*}
This relation can be used to eliminate, for example, the volume fraction perturbation $\hat\beta$:
\begin{equation*}
  \hat\beta = \frac{b(z)}{a(z)}\hat q +
  \Bigl(\od{\underline{\alpha}}{z}
   + \rho_0 g \frac{b(z)}{a(z)}\Bigr)\frac{\hat w}{i\omega}.
\end{equation*}
Thus, the system governing the behaviour of the perturbations $\hat w$ and $\hat q$ is given by the two equations
\begin{eqnarray}
  \od{\hat w}{z} + i\Bigl(\frac{|\k|^2}{\rho_0\omega}-\frac{\omega}{a(z)}\Bigr)\hat q - \frac{\rho_0 g}{a(z)} \hat w &=& 0, \label{eq:1} \\
  \frac{1}{\rho_0}\od{\hat q}{z} + g \Bigl(d(z) + c(z)\frac{b(z)}{a(z)}\Bigr)\hat q
  +\left[\omega^2 + g c(z)\Bigl(\od{\underline{\alpha}}{z} + \rho_0 g \frac{b(z)}{a(z)}\Bigr)\right]\frac{\hat w}{i\omega} &=& 0 \label{eq:2}.
\end{eqnarray}
Note that this analysis is consistent with the previous one without gravity. Indeed, if one takes $g=0$ in
(\ref{eq:1})-(\ref{eq:2}) and assumes a periodic dependence in $z$ with wavenumber $k_3$, one recovers the previous
dispersion relation.

In order to find the dispersion relation in the general case, these equations must completed by boundary conditions,
for example
\begin{equation*}
  \hat w_{\rm bottom} = 0; \quad \hat w_{\rm top} = 0\,,
\end{equation*}
if the flow occurs between two solid walls.

One expects to obtain a dispersion relation as a solvability condition for the second-order boundary value problem. Unfortunately, at this stage, we cannot go much further, except for the particular case where ${gz}/{(c_s^\pm)^2} \ll 1$. Then one can perform an asymptotic expansion in this small parameter. As a result the coefficients are polynomials in $z$. Restricting to the leading order terms yields a system of two ordinary differential equations with coefficients that are affine in $z$. The solution can be obtained in terms of Airy's wave functions Ai and Bi.

\subsection{Pure fluid limit}

We show now that the two-fluid model degenerates into the classical water-wave equations in the limit of an
interface separating two pure fluids.
Consider the case where $\alpha$ is either $1$ or $-1$. More precisely let
\begin{equation}\label{alpha_i}
    \alpha := 1 - 2{\mathcal H}(z-\eta(\x,t))\,, \quad \x = (x_1,x_2)
\end{equation}
where ${\mathcal H}$ is the Heaviside step function. Physically this substitution means that we consider two pure fluids separated by an interface. It follows that
$$
  \alpha^+ \alpha^- = 0\,, \quad 1-\alpha^2 = 0\,.
$$

Substituting the expression (\ref{alpha_i}) into the equation (\ref{eq:alpha}) gives
\begin{equation*}
    \eta_t + \u_h\cdot\nabla_h\eta = w\,,
\end{equation*}
where $\u_h = (u_1,u_2)$ and $\nabla_h = (\partial_{x_1}, \partial_{x_2})$.

This equation simply states that there is no mass flux across the interface. Incidentally this is no longer true in the case of shock waves.  Integrating the conservation of momentum equation (\ref{eq:momentumphys}) inside a volume moving with the
flow and enclosing the interface and using the fact there is no mass flux across the interface simply leads to the fact that there is no pressure jump across the interface. In other words, the pressure is continuous across the interface. Integrating the entropy equation
inside the same volume enclosing the interface and using the fact there is no mass flux across the interface does not lead
to any new information.

One can now write Eqs (\ref{eq:massphys2})--(\ref{eq:energyphys}) in each fluid, either in the conservative form
\begin{eqnarray}
(\rho^\pm)_t + \div(\rho^\pm \u^\pm) &=& 0\,, \\
(\rho^\pm\u^\pm)_t + \div(\rho^\pm \u^\pm \otimes \u^\pm)+ \nabla p^\pm &=& \rho^\pm\g\,, \\
(\rho^\pm s^\pm)_t  + \div (\rho^\pm s^\pm \u^\pm) = 0\,,
\end{eqnarray}
or in the more classical form
\begin{eqnarray}\label{linclass1}
	\rho^\pm_t+(\u^\pm\cdot\nabla)\rho^\pm + \rho^\pm \div\u^\pm & = & 0\,, \\ \label{linclass2}
	\u^\pm_t+(\u^\pm\cdot\nabla)\u^\pm + \frac{\nabla p^\pm}{\rho^\pm} &=& \g\,, \\ \label{linclass3}
	s^\pm_t+\u^\pm\cdot\nabla s^\pm &=& 0\,.
\end{eqnarray}
In these two systems, the superscripts $+$ and $-$ are used for the heavy fluid (below the interface) and the light fluid (above the interface) respectively.

The system of equations we derived is nothing else than the system of a discontinuous two-fluid system with an interface
located at $z=\eta(\x,t)$. Along the interface, one has the kinematic and dynamic boundary conditions
\begin{eqnarray}
	\eta_t + \u_h^\pm\cdot\nabla_h\eta &=& w^\pm \\
	p^- &=& p^+ 
\end{eqnarray}

This simple computation shows an interesting property of our model: it automatically degenerates into a discontinuous two-fluid system where two pure compressible phases are separated by an interface. In Appendix \ref{pure}, we derive
the dispersion relation for this limit. We choose the approximate rest state 
$$
\underline{\eta}= 0, \;\;
\underline{\rho}^\pm = \rho_0^\pm, \;\;
\underline{\u}^\pm = \vec{0}, \;\;
\underline{p}^\pm = p_0 - \rho_0^\pm gz, \;\;
\underline{s}^\pm = s_0^\pm,
$$
and assume that the fluid domain is bounded by two horizontal walls located at $z=\mp \alpha_0^\pm D$ ($D$ is the total
depth of the domain). 
Let $ \theta = {\rho_0^-}/{\rho_0^+}$ and introduce
\begin{equation}
	S_\omega^+ = \sqrt{1-\frac{\omega^2}{|\k|^2\left(c_s^+\right)^2}}, \quad
	T_\omega^- = \sqrt{\frac{\omega^2}{|\k|^2\left(c_s^-\right)^2}-1}, \quad
	S_\omega^- = \sqrt{1-\frac{\omega^2}{|\k|^2\left(c_s^-\right)^2}}.
\end{equation}
There is a distinction between three cases. When $c_s^-|\k| < \omega < c_s^+ |\k|$, one finds the dispersion relation
(\ref{acou}), which is reproduced here:
\begin{multline}
	\frac{\omega^2}{g|\k|}\left(
	T_\omega^-\tan(T_\omega^-\alpha_0^- |\k|D) - \theta S_\omega^+\tanh(S_\omega^+\alpha_0^+ |\k|D)
	\right) =\\=
	(1-\theta)S_\omega^+T_\omega^-\tan(T_\omega^-\alpha_0^- |\k|D)\tanh(S_\omega^+\alpha_0^+ |\k|D).
\label{acou_main}
\end{multline}

Neglecting the effects due to gravity, the dispersion relation (\ref{acou_main}) becomes
\begin{equation}\label{eq:Tomega}
	T_\omega^-\tan(T_\omega^-\alpha_0^- |\k|D) = \theta S_\omega^+\tanh(S_\omega^+\alpha_0^+ |\k|D).
\end{equation}
There is an infinite number of solutions because of the presence of the tangent term $\tan(T_\omega^-\alpha_0^- |\k|D)$. Since $\theta$ is small in coastal
engineering applications, we expect $T_\omega^-\tan(T_\omega^-\alpha_0^- |\k|D)$ to be small. Then either $(T_\omega^-)^2$ is small, or $T_\omega^-\alpha_0^- |\k|D \approx n\pi$, $n\in\Z$.

After some simple calculations, one obtains for $(T_\omega^-)^2$ small
$\omega \approx c_s^-|\k|$,
and for $T_\omega^-\alpha_0^- |\k|D\approx n\pi$
$$
	\omega_n = c_s^-|\k| \sqrt{1+\frac{n^2\pi^2}{(\alpha_0^-)^2 |\k|^2 D^2}}, \quad \mbox{with} \;\;
  \lim_{|\k|D\to 0} \omega_n = \frac{n \pi c_s^-}{\alpha_0^- D}.
$$

\begin{figure}
	\centering
		\includegraphics[width=0.80\textwidth]{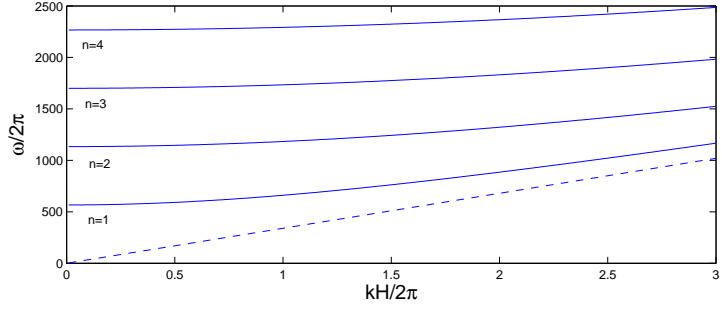}
	\caption{Dispersion relation (\ref{eq:Tomega}) for the acoustic mode.}
	\label{disp-acou}
\end{figure}

When $\omega < \min(c_s^-|\k|,c_s^+ |\k|)$, one finds the dispersion relation
(\ref{gravity}), which is reproduced here:
\begin{multline}
  \frac{\omega^2}{g|\k|}\left(
  S_\omega^-\tanh(S_\omega^-\alpha_0^- |\k|D) +
  \theta S_\omega^+\tanh(S_\omega^+\alpha_0^+ |\k|D)\right) = \\ =
  (1-\theta)S_\omega^- S_\omega^+\tanh(S_\omega^-\alpha_0^-|\k|D) \tanh(S_\omega^+\alpha_0^+|\k|D).
  \label{gravity_main}
\end{multline}
In the incompressible limit, the speeds of sound $c_s^\pm$ go to infinity,
$S_\omega^\pm \to 1$ and one recovers the classical dispersion relation for interfacial waves, namely
$$
  \frac{\omega^2}{g|\k|}\left(
  \tanh(\alpha_0^- |\k|D) +
  \theta \tanh(\alpha_0^+ |\k|D)\right) =
  (1-\theta) \tanh(\alpha_0^-|\k|D) \tanh(\alpha_0^+|\k|D).
$$

Finally, when $\omega > \max(c_s^+|\k|,c_s^- |\k|)$, one finds the dispersion relation (\ref{acoubis}), which is reproduced here:
\begin{multline}
	\frac{\omega^2}{g|\k|}\left(
	T_\omega^-\tan(T_\omega^-\alpha_0^- |\k|D) + \theta T_\omega^+\tan(T_\omega^+\alpha_0^+ |\k|D)
	\right) =\\=
	-(1-\theta)T_\omega^+T_\omega^-\tan(T_\omega^-\alpha_0^- |\k|D)\tan(T_\omega^+\alpha_0^+ |\k|D).
\label{acoubis_main}
\end{multline}
Neglecting the effects due to gravity, the dispersion relation (\ref{acoubis_main}) becomes
\begin{equation}
	T_\omega^-\tan(T_\omega^-\alpha_0^- |\k|D) + \theta T_\omega^+\tan(T_\omega^+\alpha_0^+ |\k|D) = 0.
\end{equation}
There is an infinite number of solutions because of the presence of the tangent term. Again, since $\theta$ is small in coastal
engineering applications, we expect $T_\omega^-\tan(T_\omega^-\alpha_0^- |\k|D)$ to be small. Then we are back to the first
case. In this third case, if we were not prescribing boundary conditions at the bottom and at the top, we could look for
perturbations which are periodic in all three directions with wavenumber $k_3^\pm$ in the $z-$direction. Equation (\ref{ode_pressure}) then gives
$$ \omega^2 = (c_s^\pm)^2 (|\k|^2 + (k_3^\pm)^2). $$
In other words, there is a relationship between the two vertical wavenumbers. This relationship is
$$ 1 + \left(\frac{k_3^+}{|\k|}\right)^2 = \left(\frac{c_s^-}{c_s^+}\right)^2 \left[ 1 + \left(\frac{k_3^-}{|\k|}\right)^2\,. \right] $$
It is reminiscent of Snell's law, which describes the relationship between the angles of incidence and refraction, when referring to waves passing through a boundary between two different isotropic media.

\section{A finite-volume discretization of the model}
\label{modelnum}

Here we describe the discretization of the model (\ref{eq:massphys})--(\ref{eq:energyphys}) by a standard cell-centered finite volume method. The computational domain $\Omega\subset\R^d$ is triangulated into a set of control volumes: $\Omega=\cup_{K\in\T} K$.
We start by integrating equation (\ref{syst_abst}) on $K$:
\begin{equation}\label{eq:conservlaw}
	\od{}{t}\int_{K} \w \;d\Omega + \sum_{L\in\N(K)}\int_{K\cap L}\F(\w)\cdot\n_{KL} \;d\sigma
  	  = \int_{K}\S(\w) \;d\Omega\,,
\end{equation}
where $\n_{KL}$ denotes the unit normal vector on $K\cap L$ pointing into $L$ and $
  \N(K) = \set{L\in\T: \area(K\cap L) \neq 0}\,.
$
Then, setting
\begin{equation*}
 \w_K(t) := \frac{1}{\vol(K)}\int_{K} \w(\x,t) \;d\Omega \;,
\end{equation*}
we approximate (\ref{eq:conservlaw}) by
\begin{equation}\label{a_resoudre}
	\od{\w_K}{t} + \sum_{L\in\N(K)} \frac{\area(L\cap K)}{\vol(K)} \Phi(\w_K, \w_L; \n_{KL}) =  \S(\w_K)\;,
\end{equation}
where the numerical flux $$\Phi(\w_K, \w_L; \n_{KL}) \approx\frac{1}{\area(L\cap K)}\int_{K\cap L}\F(\w)\cdot\n_{KL} \;d\sigma\,,$$ is explicitly computed by the FVCF formula of Ghidaglia {\it et al.} \cite{Ghidaglia2001}:
\begin{equation}\label{CFFV}
\Phi(\vv, \w; n)=\frac{\F(\vv) \cdot\n+\F(\w) \cdot\n}{2}-\sgn(\A_n(\mu(\vv,\w)))\frac{\F(\w) \cdot\n-\F(\vv) \cdot\n}{2}\,,
\end{equation}
where the Jacobian matrix $\A_n(\mu)$ is defined in (\ref{mat_jacob}), $\mu(\vv,\w)$ is an arbitrary mean between $\vv$ and $\w$ and $\sgn(M)$ is the matrix whose eigenvectors are those of $M$ but whose eigenvalues are the signs of the eigenvalues of $M$.

So far we have not discussed the case where a control volume $K$ meets the boundary of $\Omega$. Here we shall only consider the case where this boundary is a wall and from the numerical point of view, we only need to find the normal flux $\F\cdot\n$. Since $
  \u(\x,t)\cdot\n = 0$ for $\x\in\partial\Omega\,,
$ we have
\begin{equation*}
  \left.(\F\cdot\n)\right|_{\x\in\partial\Omega} = (0, 0, p_b n_1, p_b n_2, 0), \quad p_b := \left.p\right|_{\x\in\partial\Omega}\,,
\end{equation*}
and following Ghidaglia and Pascal \cite{Ghidaglia2005}, we can take  $p_b = p + \rho u_n c_s,$
where the right-hand side is evaluated in the control volume $K$.

\begin{rem}
 In order to turn (\ref{a_resoudre}) into a numerical algorithm, we must at least perform time discretization and give an expression for $\mu(\vv,\w)$. Since this matter is standard, we do not give the details here but instead refer to Dutykh \cite{DUT07}. Let us also notice that formula (\ref{a_resoudre}) leads to a first-order scheme but in fact we use a MUSCL technique to achieve better accuracy in space \cite{leer2006}.
\end{rem}


\section{Numerical simulations}
\label{modelres}


\subsection{Basic tests}

In order to check the accuracy of our second-order scheme, we first solve numerically the scalar linear advection equation
\begin{equation*}
  \pd{v}{t} + \a\cdot\grad v = 0, \quad \a\in\R^2\,,
\end{equation*}
with smooth initial conditions with compact support in order to reduce the influence of boundary conditions. It is obvious that this equation will just translate the initial form in the direction $\a$. So, we have an analytical solution which can be used to quantify the error of the numerical method. On the other hand, to measure the convergence rate, we constructed a sequence of refined meshes.

\begin{figure}[htbp]
	\centering
		\includegraphics[width=0.99\textwidth]{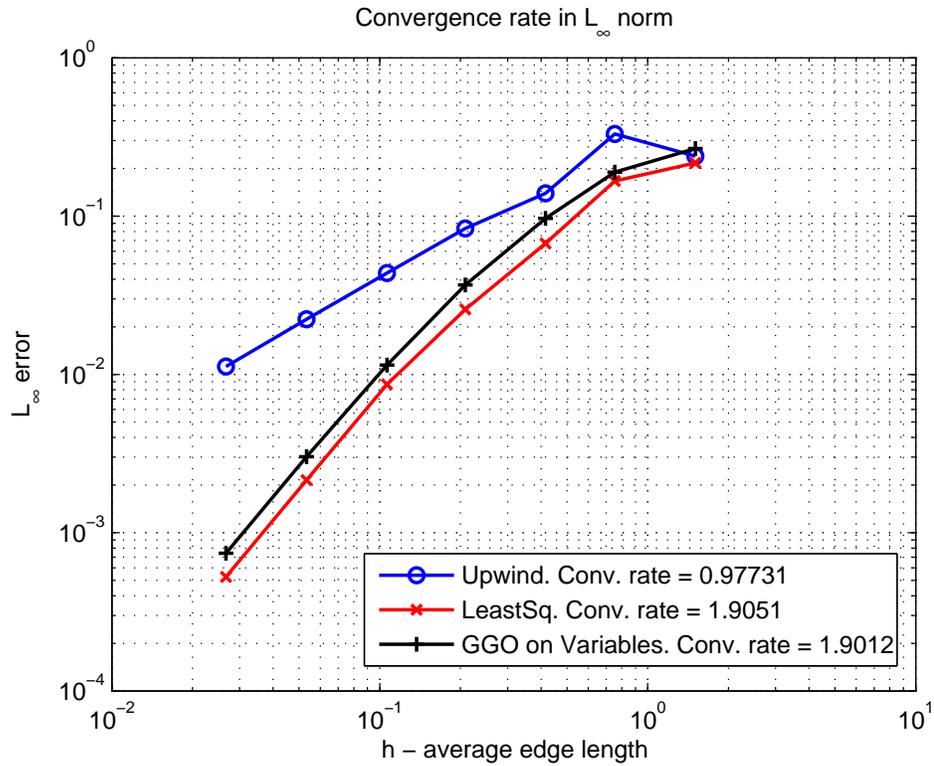}
	\caption{Numerical method error in $L_\infty$ norm.}
	\label{fig:convlinf}
\end{figure}

\figurename~\ref{fig:convlinf} shows the error of the numerical method in $L_\infty$ norm as a function of the mesh characteristic size. The slope of the curves represents an approximation to the theoretical convergence rate. On this plot, the curve with circles for the data points corresponds to the first order upwind scheme while the other two correspond to the MUSCL scheme with least-squares and Green-Gauss gradient reconstruction procedures respectively. The slope of the curve with
circles is equal approximatively to $0.97$, which means first-order convergence. The other two curves have almost the same slope equal to $1.90$, thus indicating a second-order convergence rate for the MUSCL scheme. Remark that in our implementation of the second-order scheme the least-squares reconstruction seems to give slightly more accurate results than the Green-Gauss procedure.

\begin{figure}[htbp]
	\centering
		\includegraphics[width=0.9\textwidth]{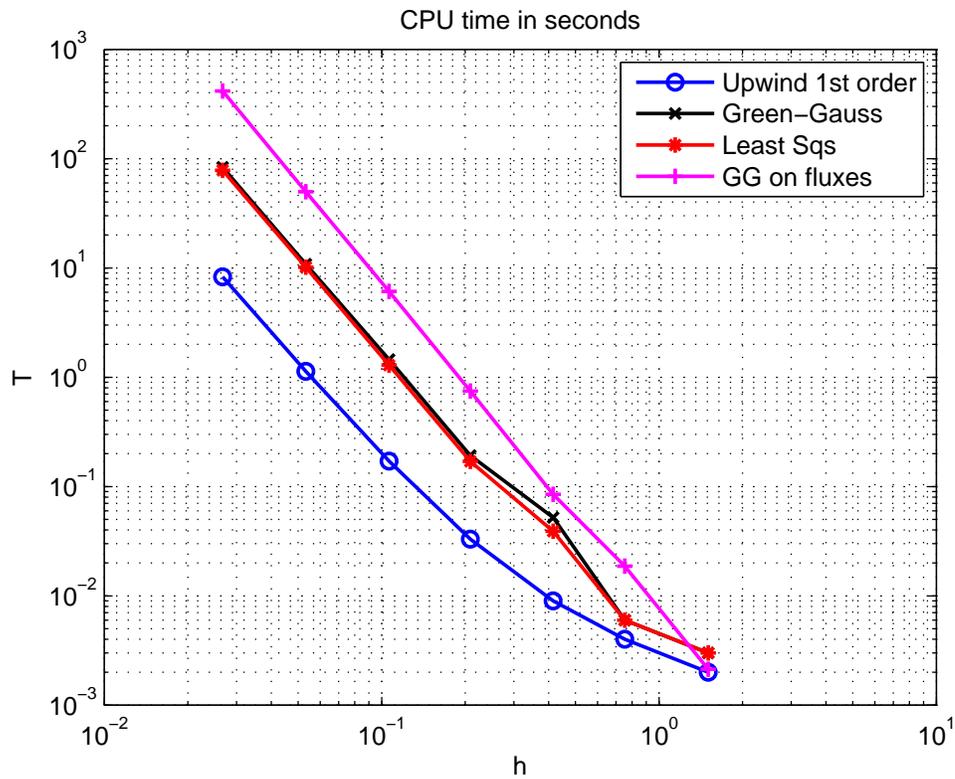}
	\caption{CPU time for different finite volume schemes.}
	\label{fig:cputime}
\end{figure}

The next figure represents the measured CPU time in seconds, again as a function of the mesh size. Obviously, this kind of data is extremely computer dependent but the qualitative behaviour is the same on all systems. On \figurename~\ref{fig:cputime} one can see that the ``fastest'' curve is that of the first-order upwind scheme. Then we have two almost superimposed curves referring to the second-order gradient reconstruction on variables. Here again one can notice that the least-squares method is slightly faster than the Green-Gauss procedure. On this figure we represented one more curve (the highest one) which corresponds to Green-Gauss gradient reconstruction on fluxes (it seems to be very natural in the context of FVCF schemes). The numerical tests show that this method is quite expensive from the computational point of view and we decided not to use it.

\begin{figure}[htbp]
	\centering
		\includegraphics[width=0.70\textwidth]{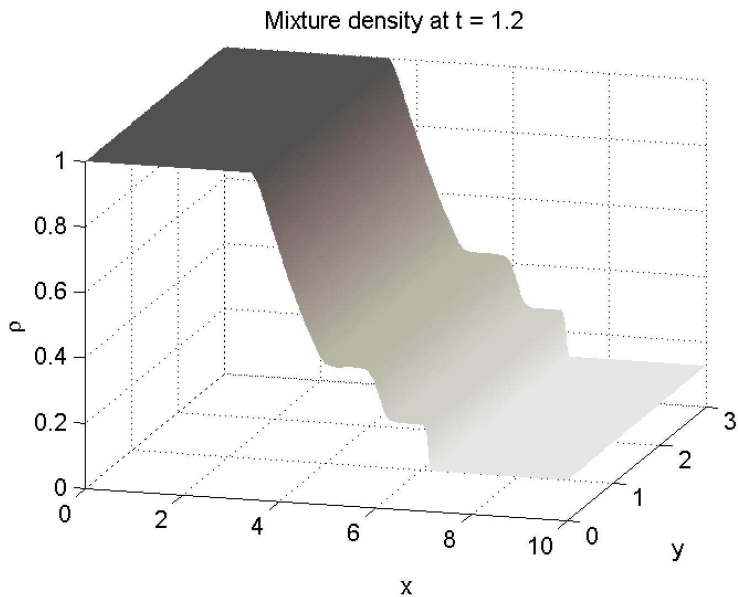}
	\caption{Shock tube of Sod: density plot.}
	\label{fig:sod}
\end{figure}

The second test is the ability of the solver to capture shocks without spurious oscillations. It is indeed the case as seen on \figurename~\ref{fig:sod}, which shows a density plot for Sod's shock tube \cite{Sod1978}.

\subsection{Falling water column}

\begin{figure}[htbp]
\centering
\psfrag{A}{$\alpha^+ = 0.9$}
\psfrag{B}{$\alpha^- = 0.1$}
\psfrag{C}{$\alpha^+ = 0.1$}
\psfrag{D}{$\alpha^- = 0.9$}
\psfrag{0}{$0$}
\psfrag{0.3}{$0.3$}
\psfrag{0.65}{$0.65$}
\psfrag{0.7}{$0.7$}
\psfrag{0.05}{$0.05$}
\psfrag{1}{$1$}
\psfrag{0.9}{$0.9$}
\psfrag{g}{$\g$}
\includegraphics[width=12cm]{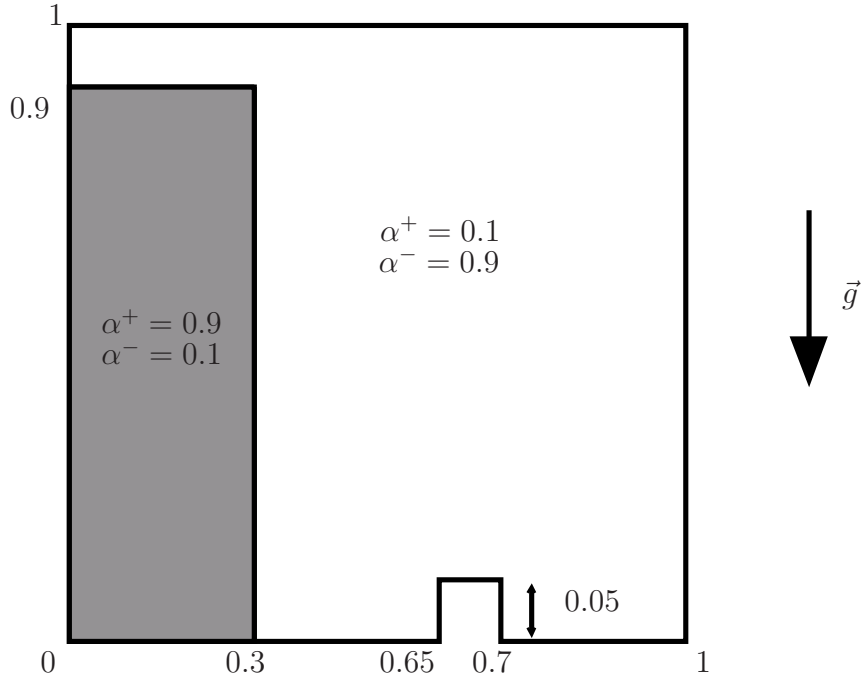}
\caption[Geometry and initial condition for falling water]{Falling water column test case. Geometry and initial condition.}
\label{fig:falling_water}
\end{figure}

\begin{figure}
	\centering
	\subfigure[$t = 0.005$]%
	{\includegraphics[width=0.46\textwidth]{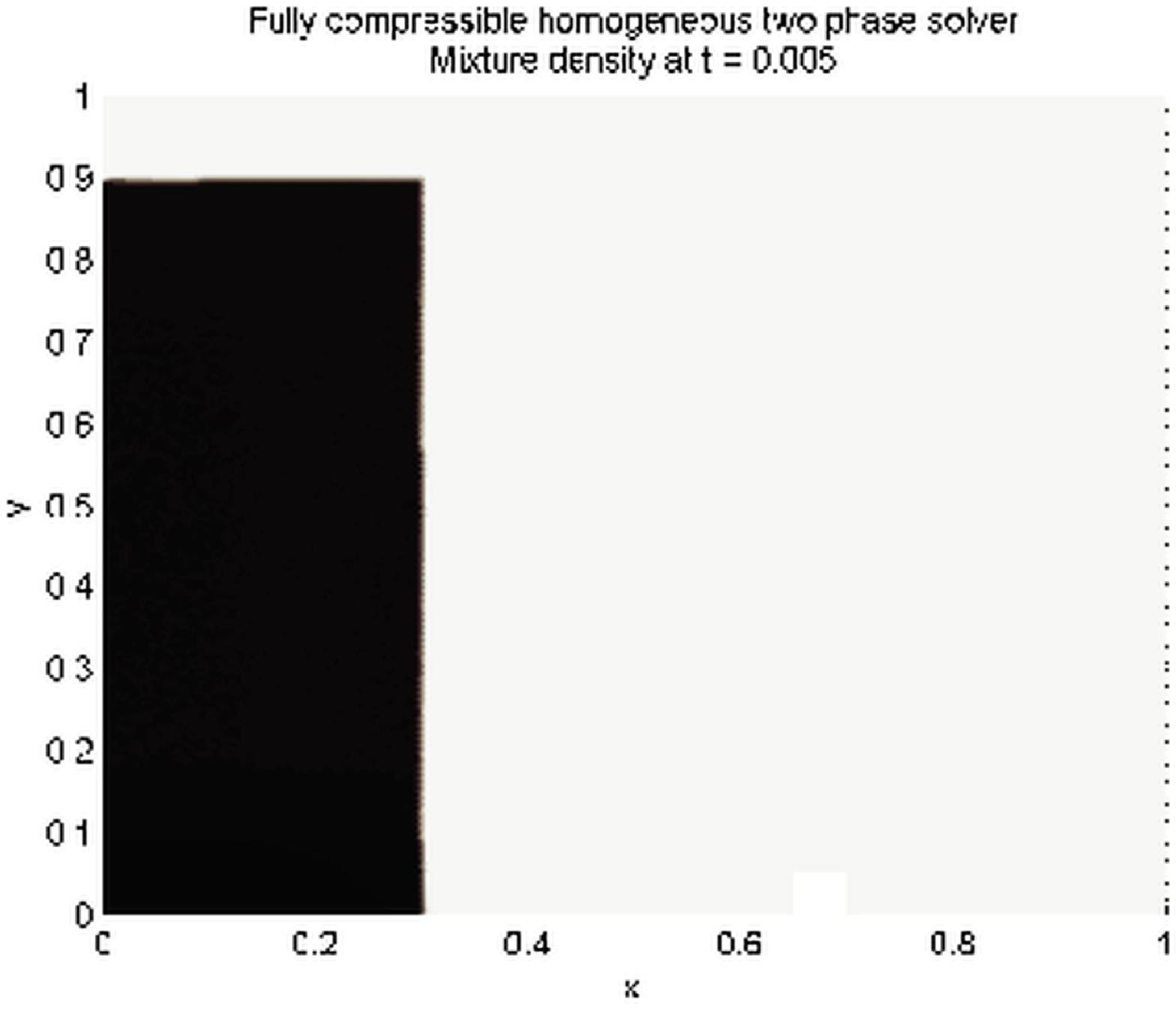}} \quad
	\subfigure[$t = 0.06$]%
	{\includegraphics[width=0.46\textwidth]{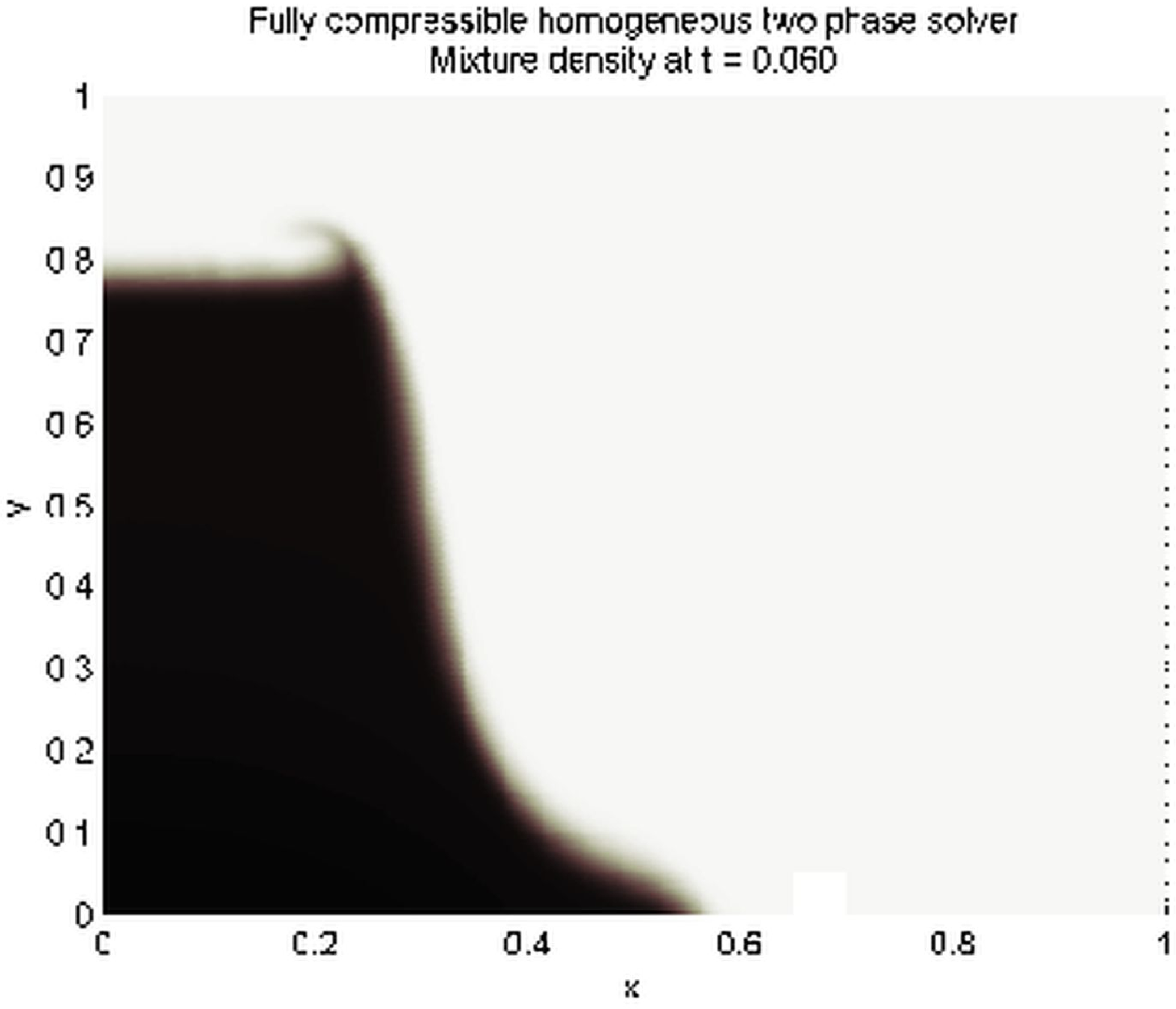}}
	\caption[The beginning of column dropping process]{Falling water column test case. Initial condition and the beginning of the column collapse.}
	\label{fig:debutSplash}
\end{figure}

\begin{figure}
	\centering
	\subfigure[$t = 0.15$]%
	{\includegraphics[width=0.46\textwidth]{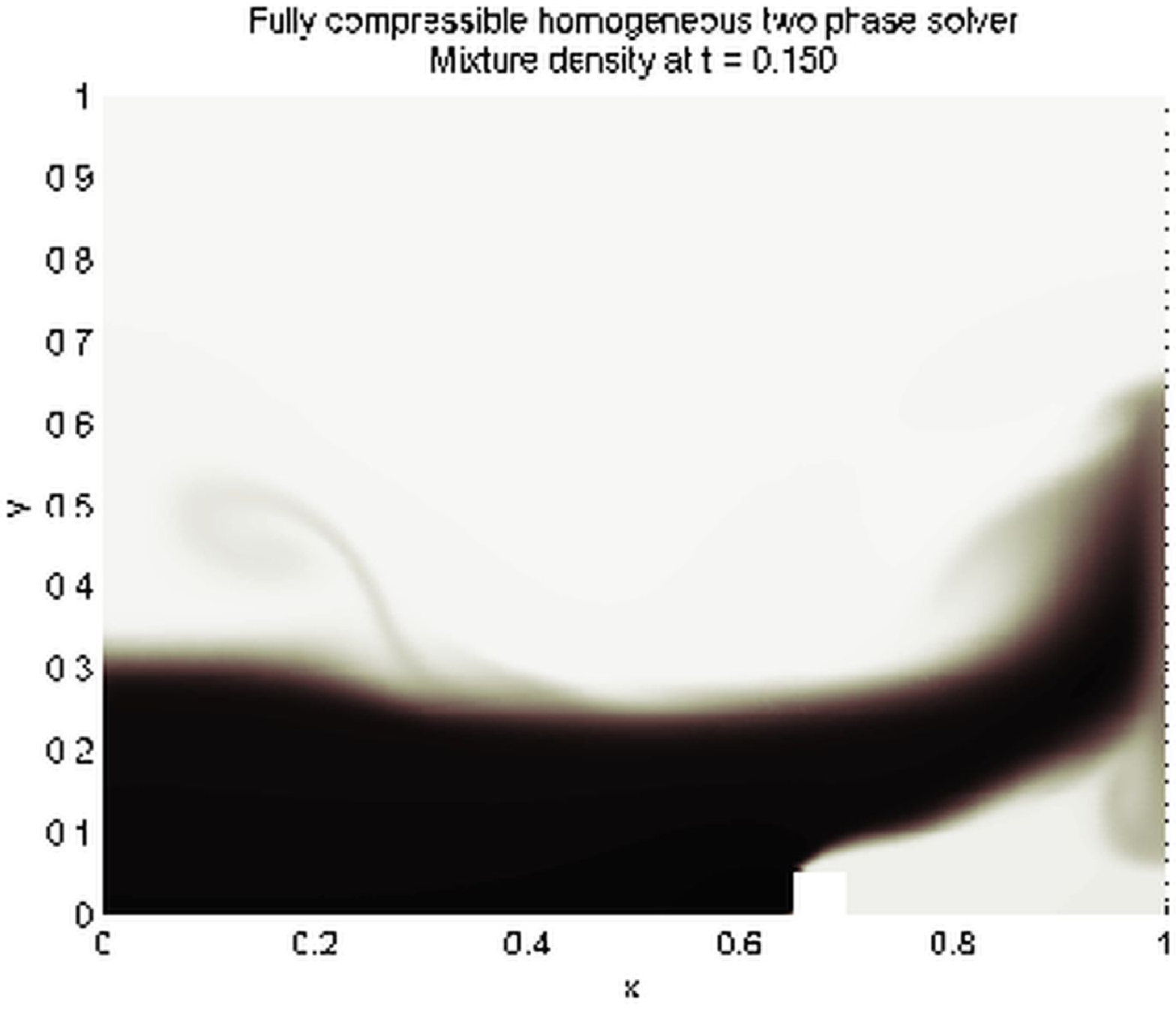}} \quad
	\subfigure[$t = 0.175$]%
	{\includegraphics[width=0.46\textwidth]{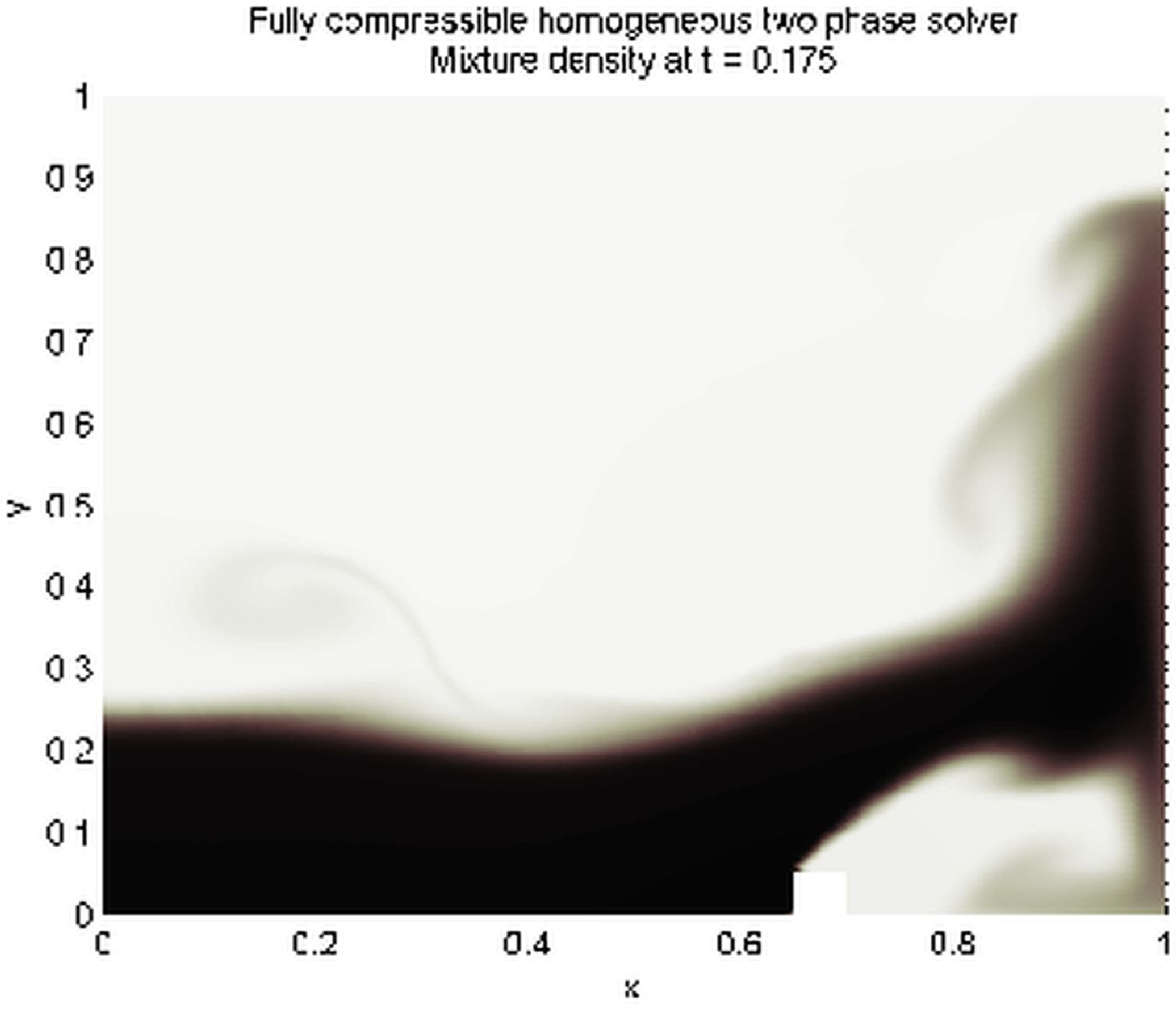}}
	\caption[Water strikes the wall - I]{Falling water column test case. Water hits the wall.}
	\label{fig:nextSplash}
\end{figure}

\begin{figure}
	\centering
	\subfigure[$t = 0.5$]%
	{\includegraphics[width=0.46\textwidth]{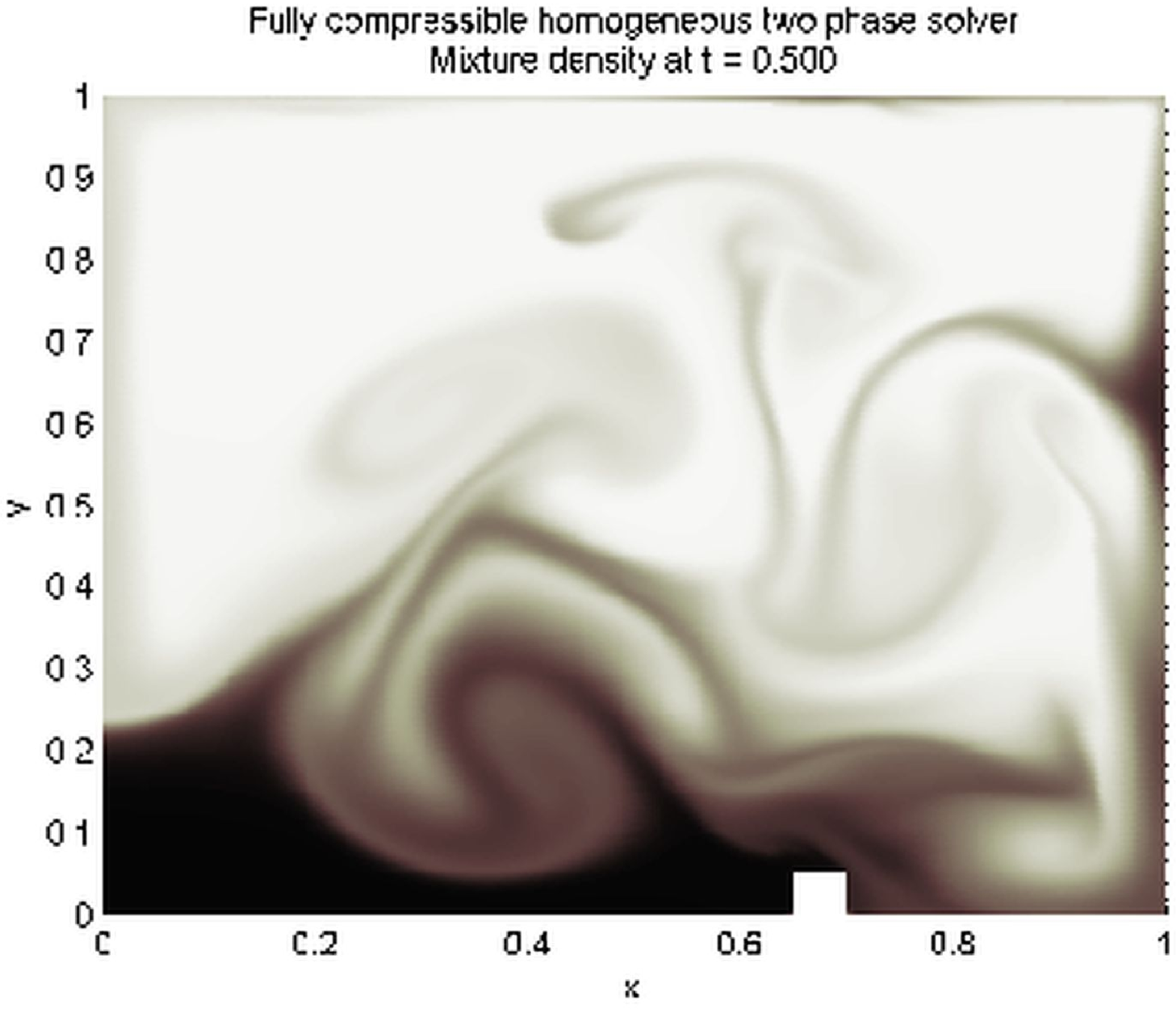}} \quad
	\subfigure[$t = 0.675$]%
	{\includegraphics[width=0.46\textwidth]{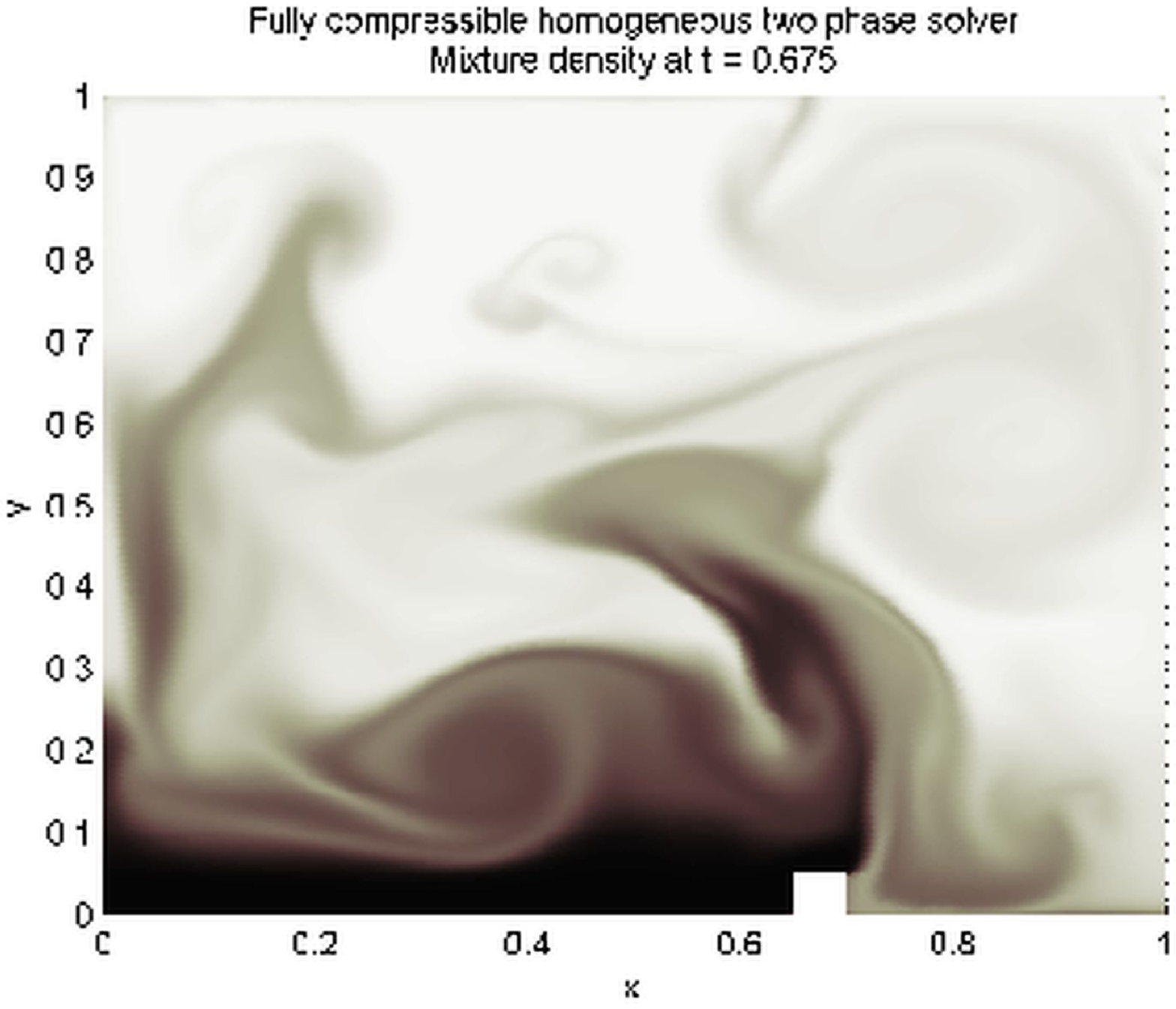}}
	\caption[Turbulent mixing process]{Falling water column test case. Turbulent mixing process.}
	\label{fig:lastSplash}
\end{figure}

\begin{figure}
	\centering
		\includegraphics[width=0.80\textwidth]{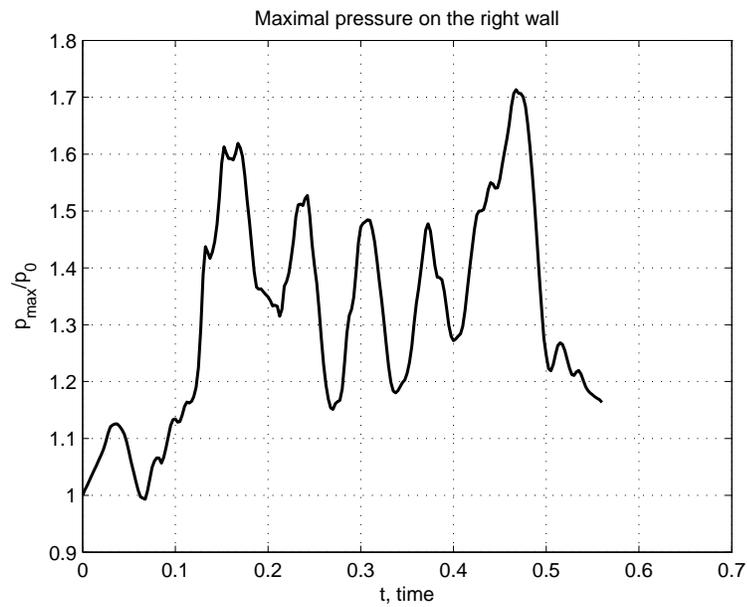}
	\caption{Maximal pressure on the right wall as a function of time. Case of a heavy gas.}
	\label{fig:wallpress1}
\end{figure}

\begin{figure}
	\centering
		\includegraphics[width=0.80\textwidth]{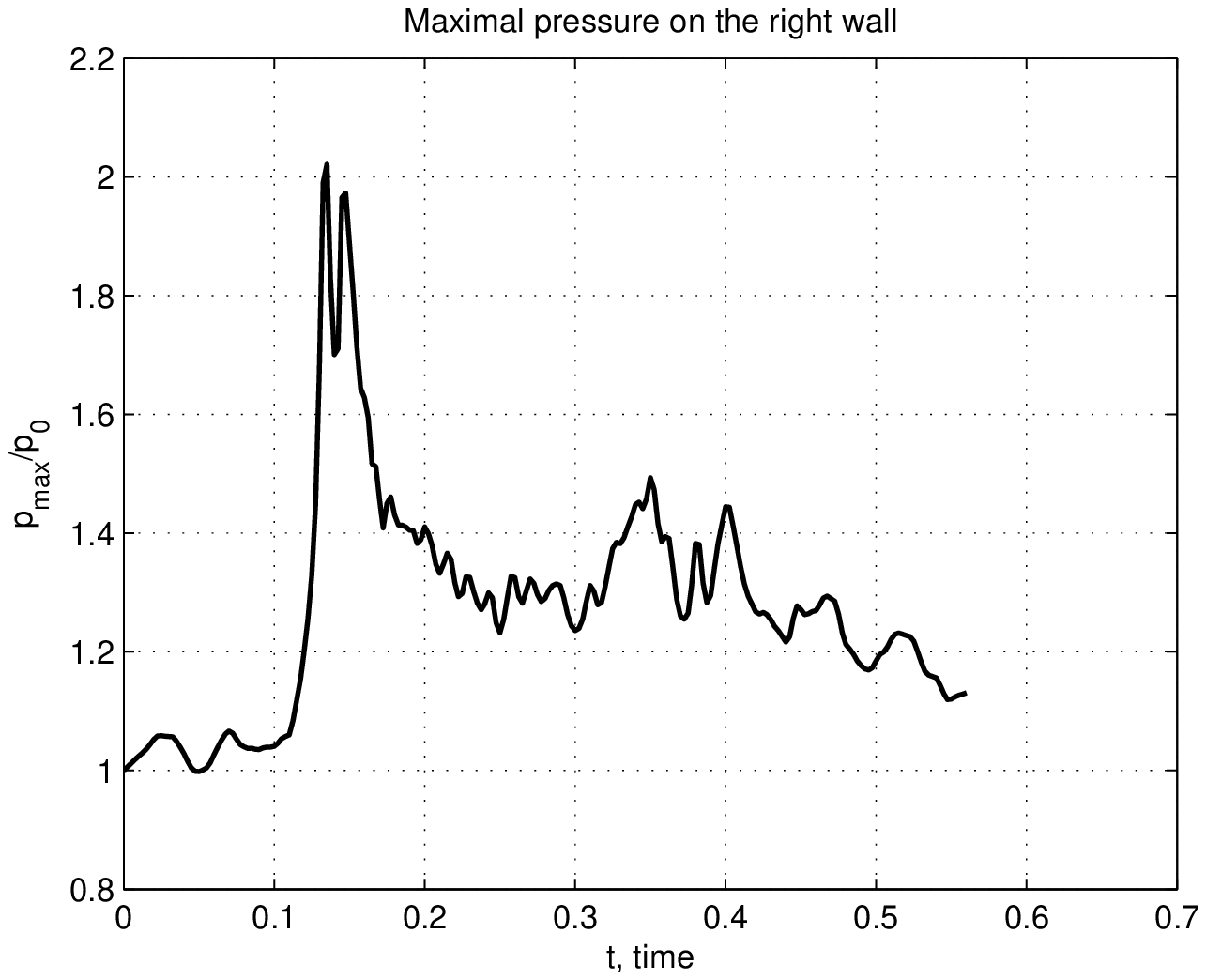}
	\caption{Maximal pressure on the right wall as a function of time. Case of a light gas.}
	\label{fig:wallpress2}
\end{figure}

The geometry and initial condition for this test case are shown on \figurename~\ref{fig:falling_water}. Initially the velocity field is taken to be zero. The values of the other parameters are given in Table \ref{tab:diphaseparams}. The mesh used in this computation contained about $108 000$ control volumes (in this case they were triangles). The results of this simulation are presented on Figures \ref{fig:debutSplash}--\ref{fig:lastSplash}.
\figurename~\ref{fig:wallpress1} shows the maximal pressure on the right wall as a function of time:
\begin{equation*}
  t \longmapsto \max_{(x,y) \in 1\times[0,1]} p(x,y,t).
\end{equation*}
We performed another computation for a mixture with $\alpha^+ = 0.05$, $\alpha^- = 0.95$. The pressure is recorded as well and plotted in \figurename~\ref{fig:wallpress2}. One can see that the peak value is higher and the impact is more localized in time.

\begin{figure}
	\centering
	\subfigure[$t = 0.1$]%
	{\includegraphics[width=0.46\textwidth]{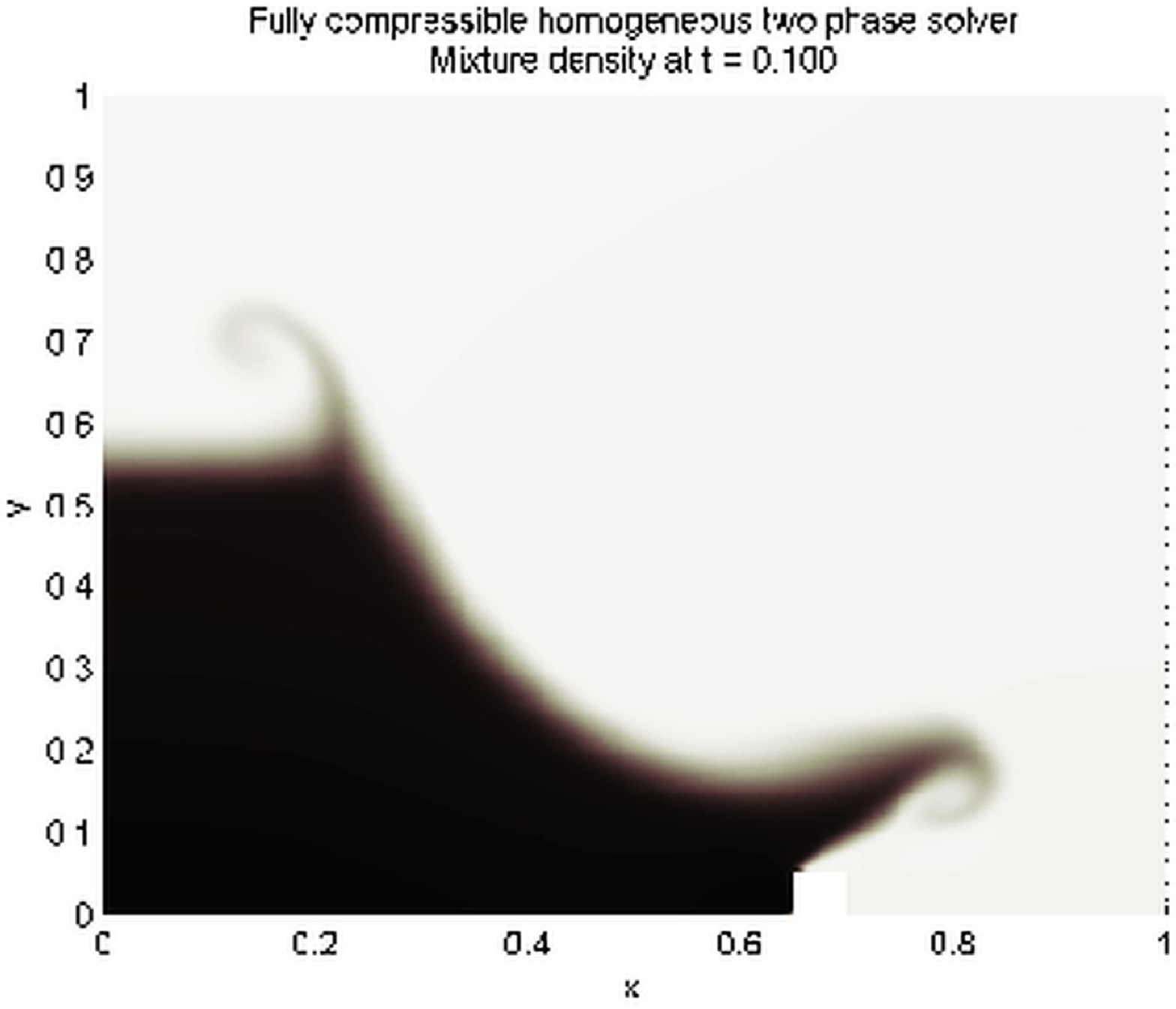}} \quad
	\subfigure[$t = 0.125$]%
	{\includegraphics[width=0.46\textwidth]{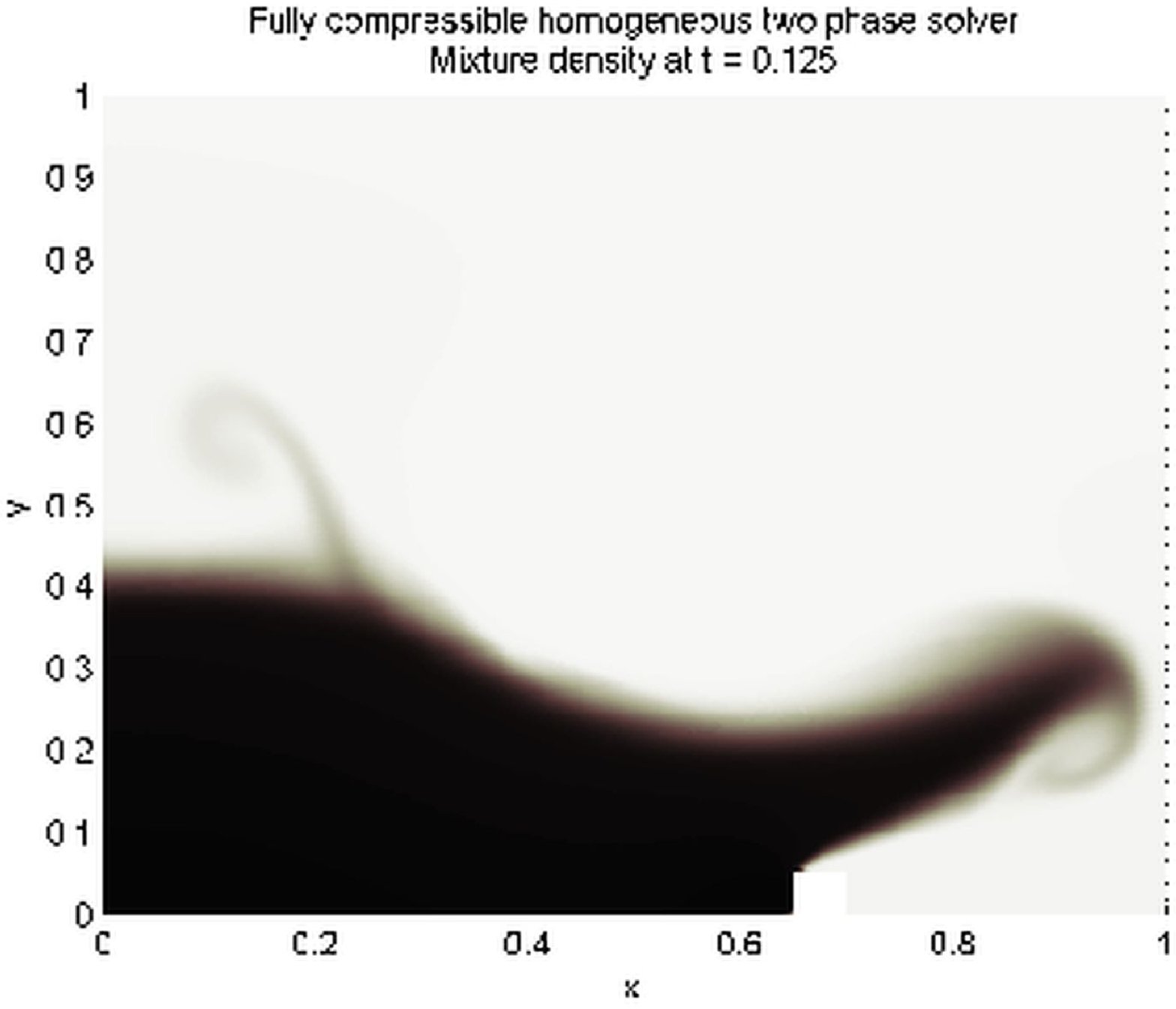}}
	\caption[Splash creation due to the interaction with step]{Falling water column test case. Splash formation due to the interaction with the step.}
\end{figure}

\begin{figure}
	\centering
	\subfigure[$t = 0.2$]%
	{\includegraphics[width=0.46\textwidth]{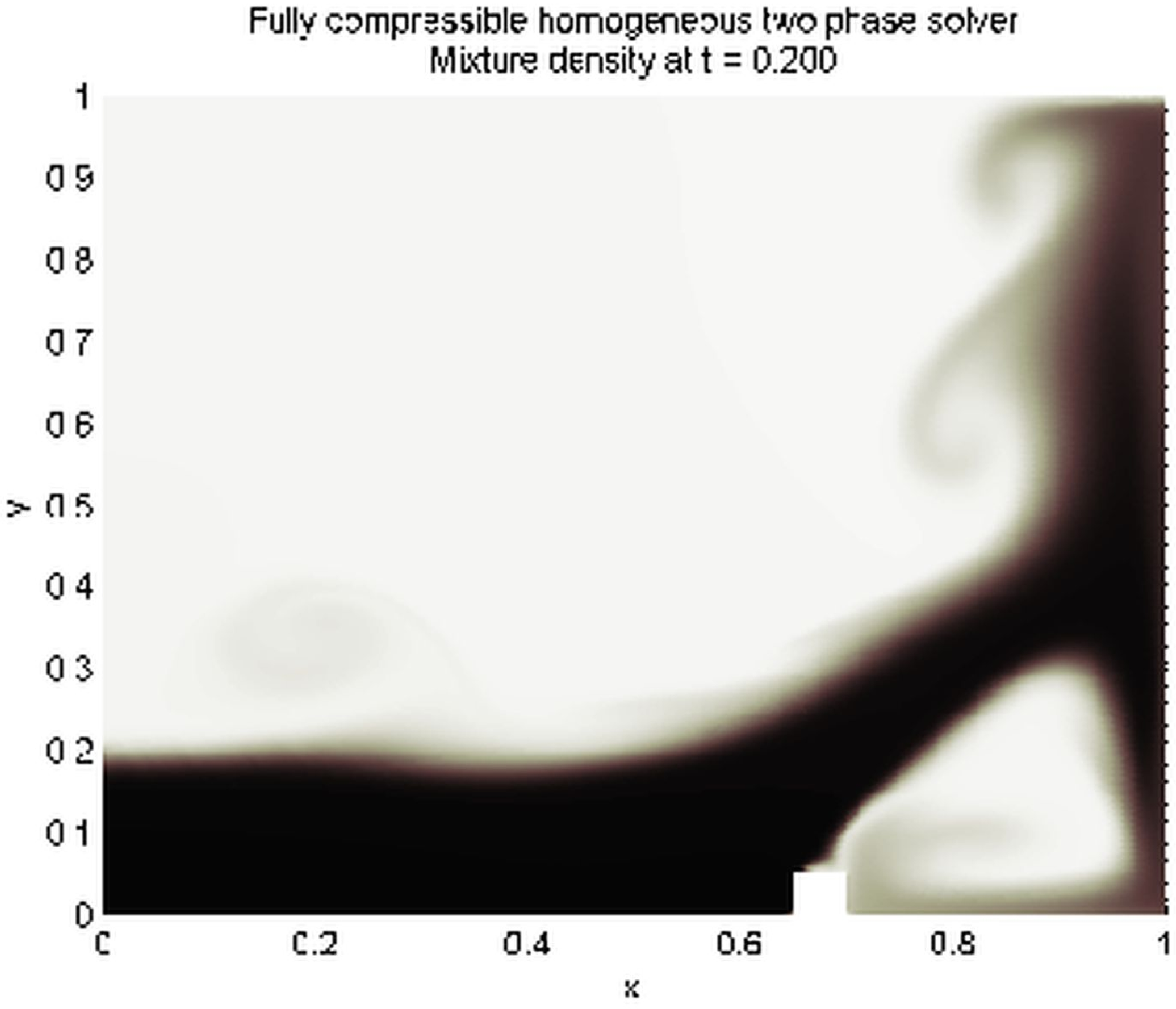}} \quad
	\subfigure[$t = 0.225$]%
	{\includegraphics[width=0.46\textwidth]{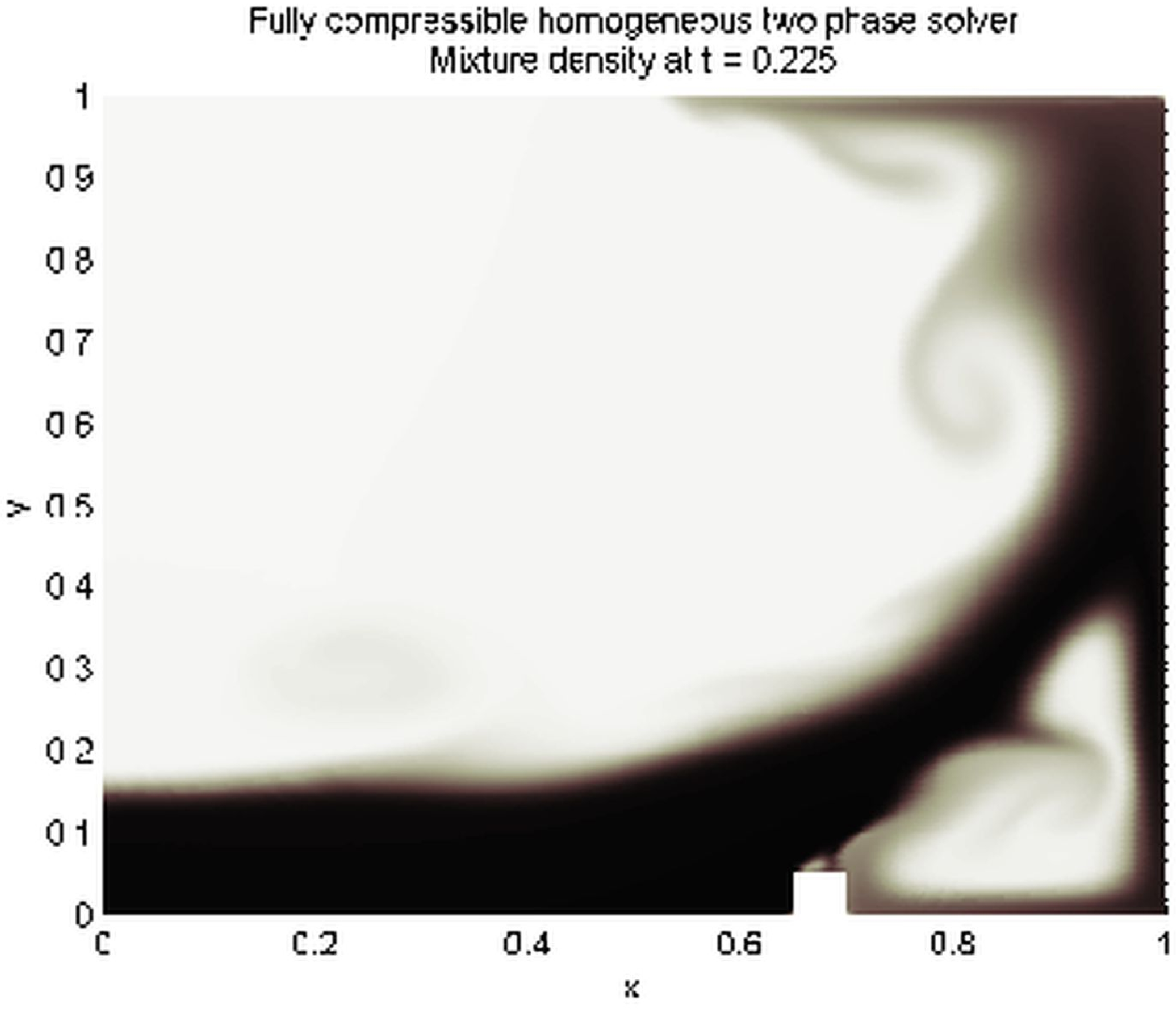}}
	\caption[Water strikes the wall - II]{Same as Fig. \ref{fig:nextSplash} at later times.}
\end{figure}

\begin{figure}
	\centering
	\subfigure[$t = 0.3$]%
	{\includegraphics[width=0.46\textwidth]{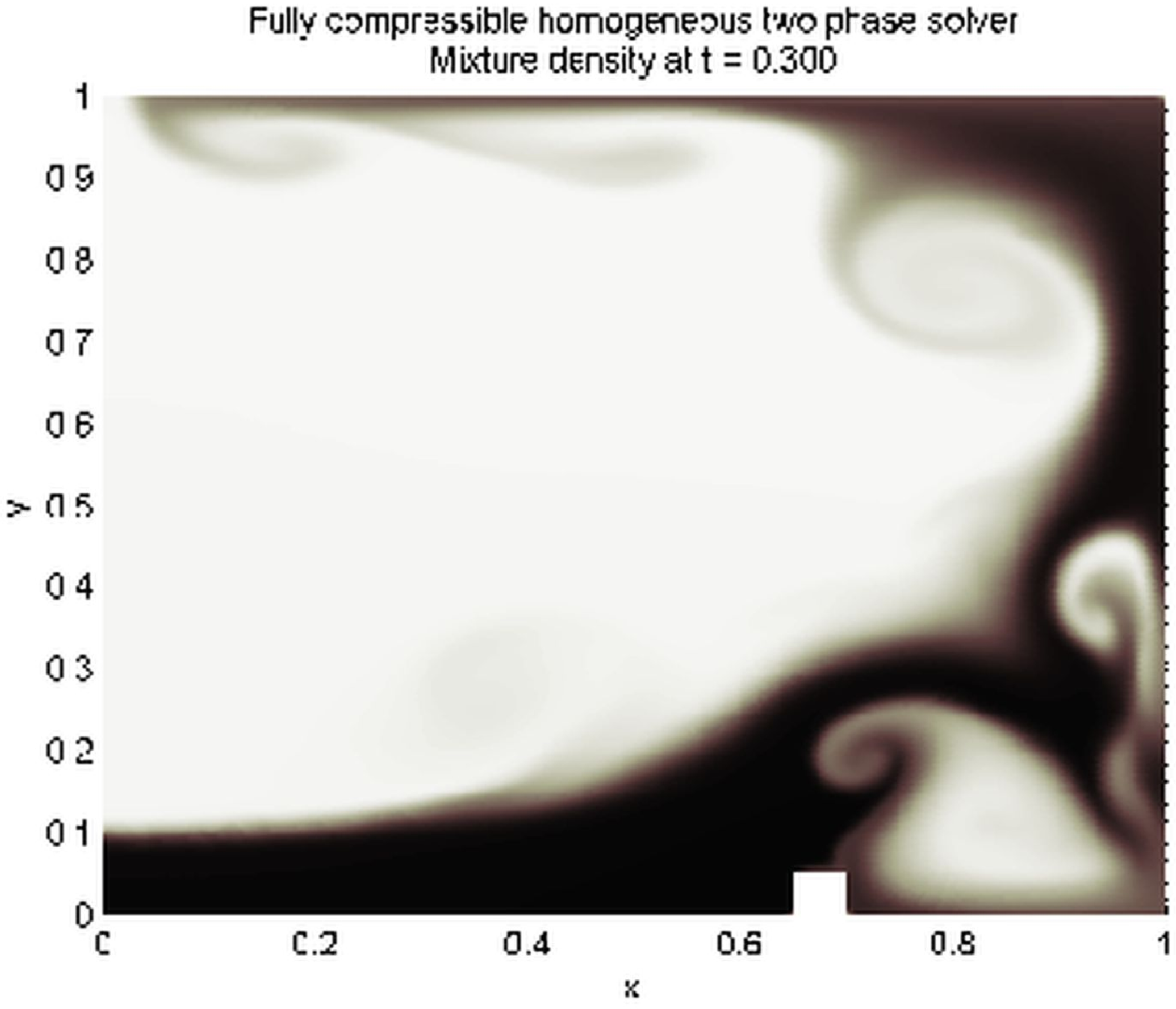}} \quad
	\subfigure[$t = 0.4$]%
	{\includegraphics[width=0.46\textwidth]{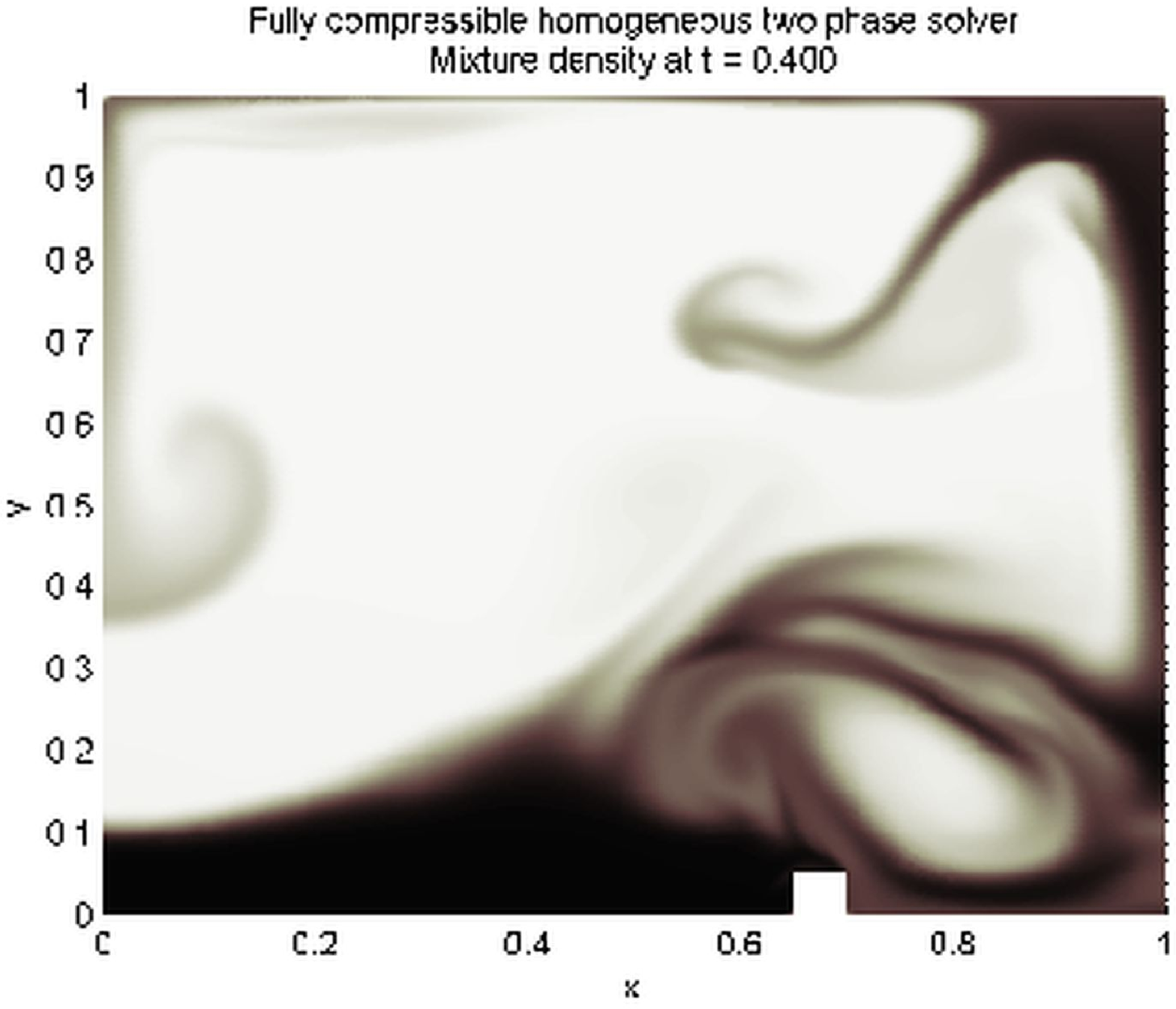}}
	\caption[Splash is climbing the wall]{Falling water column test case. The splash is climbing the wall.}
\end{figure}

\subsection{Water drop test case}

\begin{figure}[htbp]
\centering
\psfrag{A}{$\alpha^+ = 0.1$}
\psfrag{B}{$\alpha^- = 0.9$}
\psfrag{C}{$\alpha^+ = 0.9$}
\psfrag{D}{$\alpha^- = 0.1$}
\psfrag{0}{$0$}
\psfrag{0.5}{$0.5$}
\psfrag{0.7}{$0.7$}
\psfrag{1}{$1$}
\psfrag{g}{$\g$}
\psfrag{R}{$R = 0.15$}
\includegraphics[width=13cm]{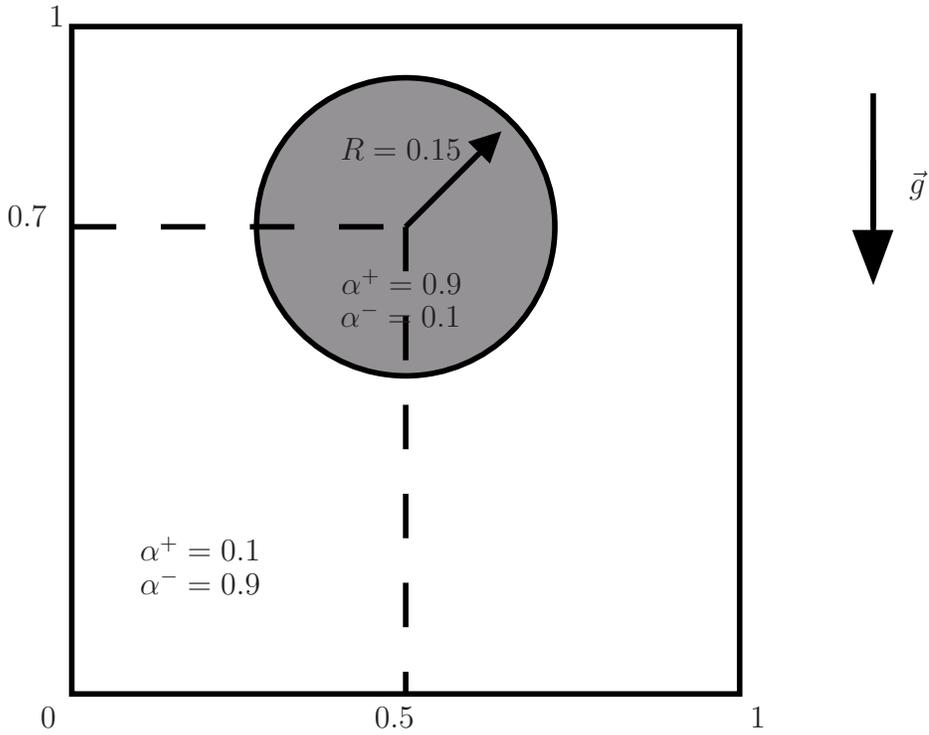}
\caption[Geometry and initial condition for water drop test case]{Geometry and initial condition for water drop test case.}
\label{fig:drop_water}
\end{figure}

\begin{figure}
	\centering
	\subfigure[$t = 0.005$]%
	{\includegraphics[width=0.46\textwidth]{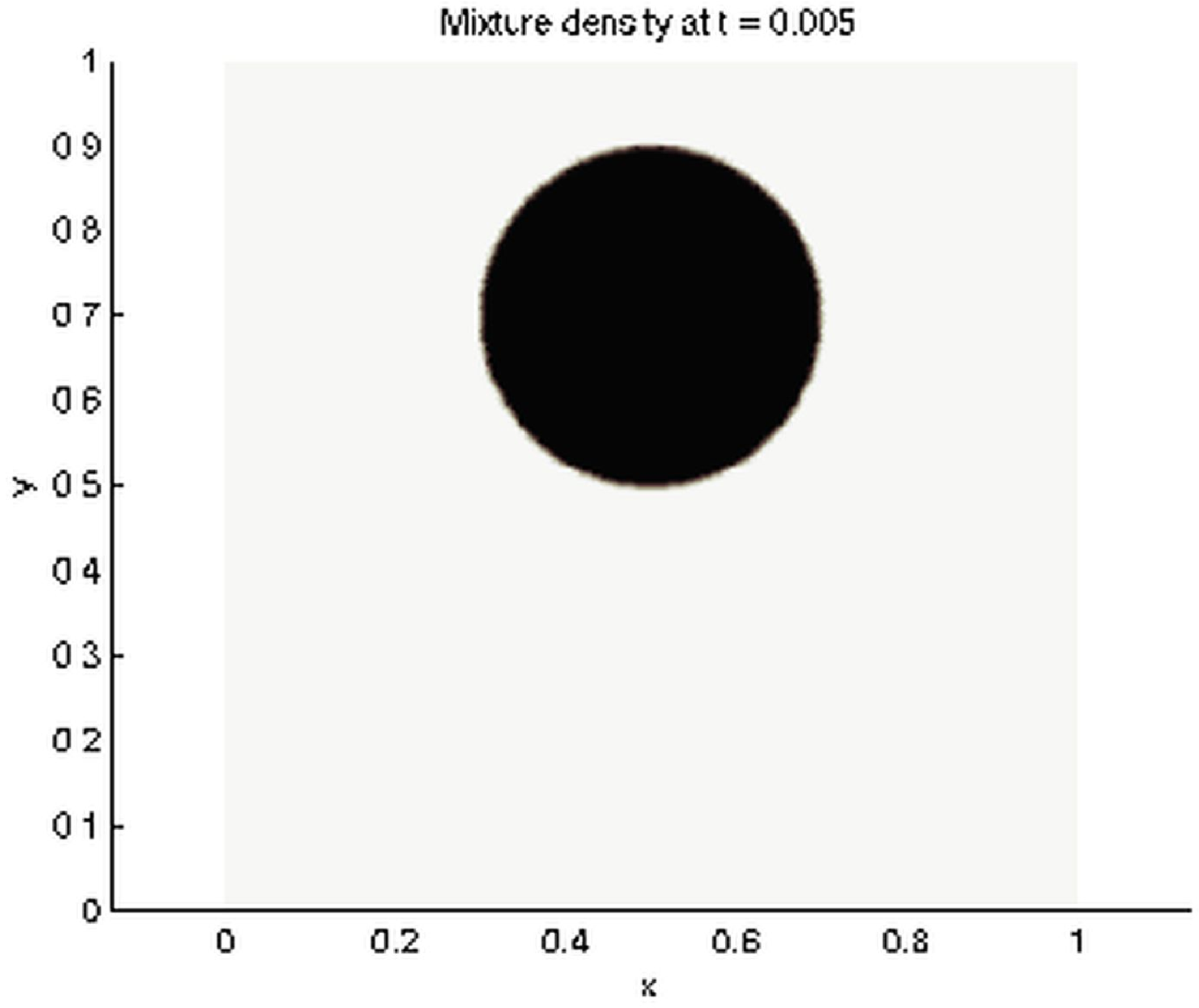}} \quad
	\subfigure[$t = 0.075$]%
	{\includegraphics[width=0.46\textwidth]{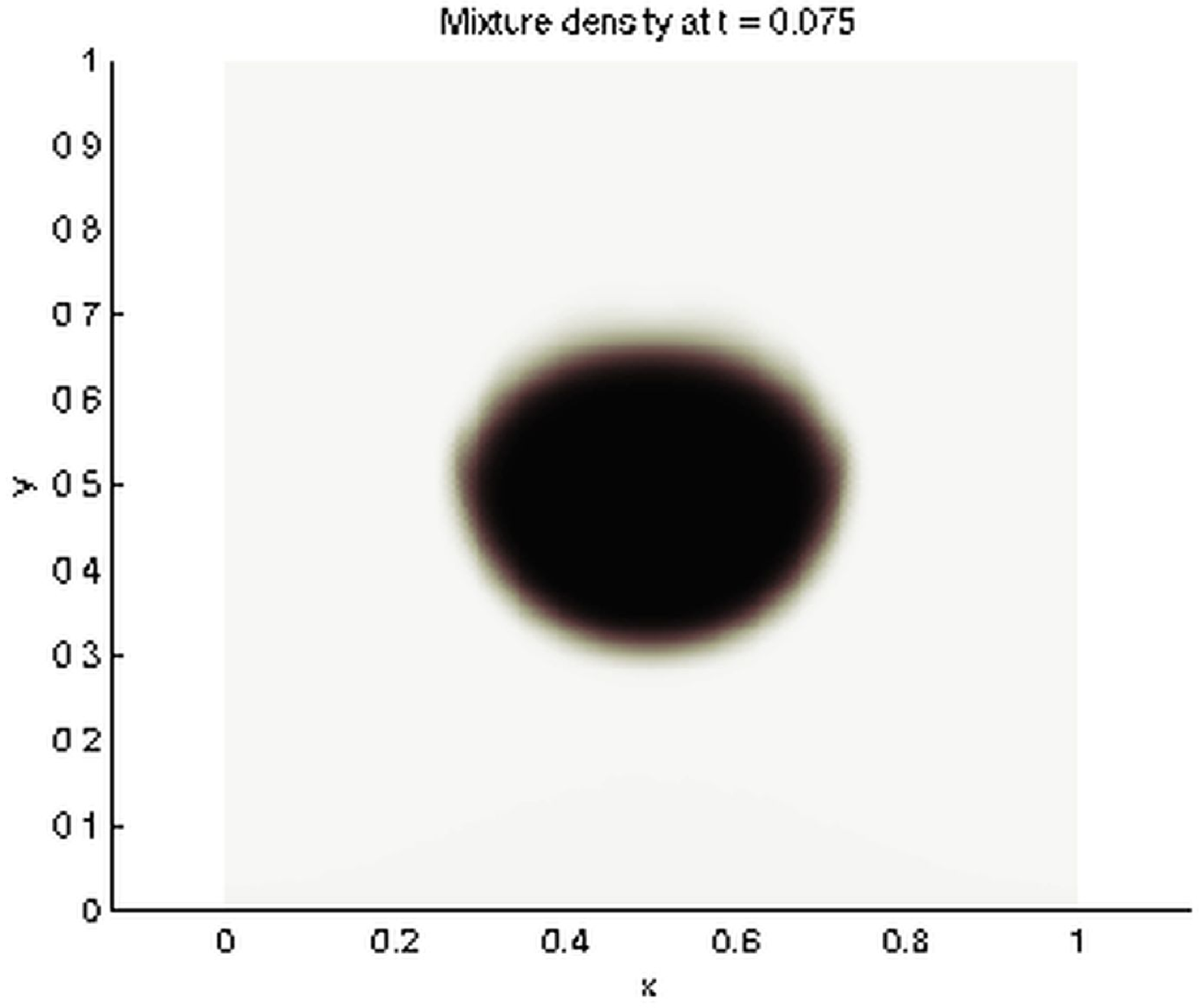}}
	\caption[Initial configuration and the beginning of the fall]{Water drop test case. Initial configuration and the beginning of the fall.}
	\label{fig:debutDrop}
\end{figure}

\begin{figure}
	\centering
	\subfigure[$t = 0.325$]%
	{\includegraphics[width=0.46\textwidth]{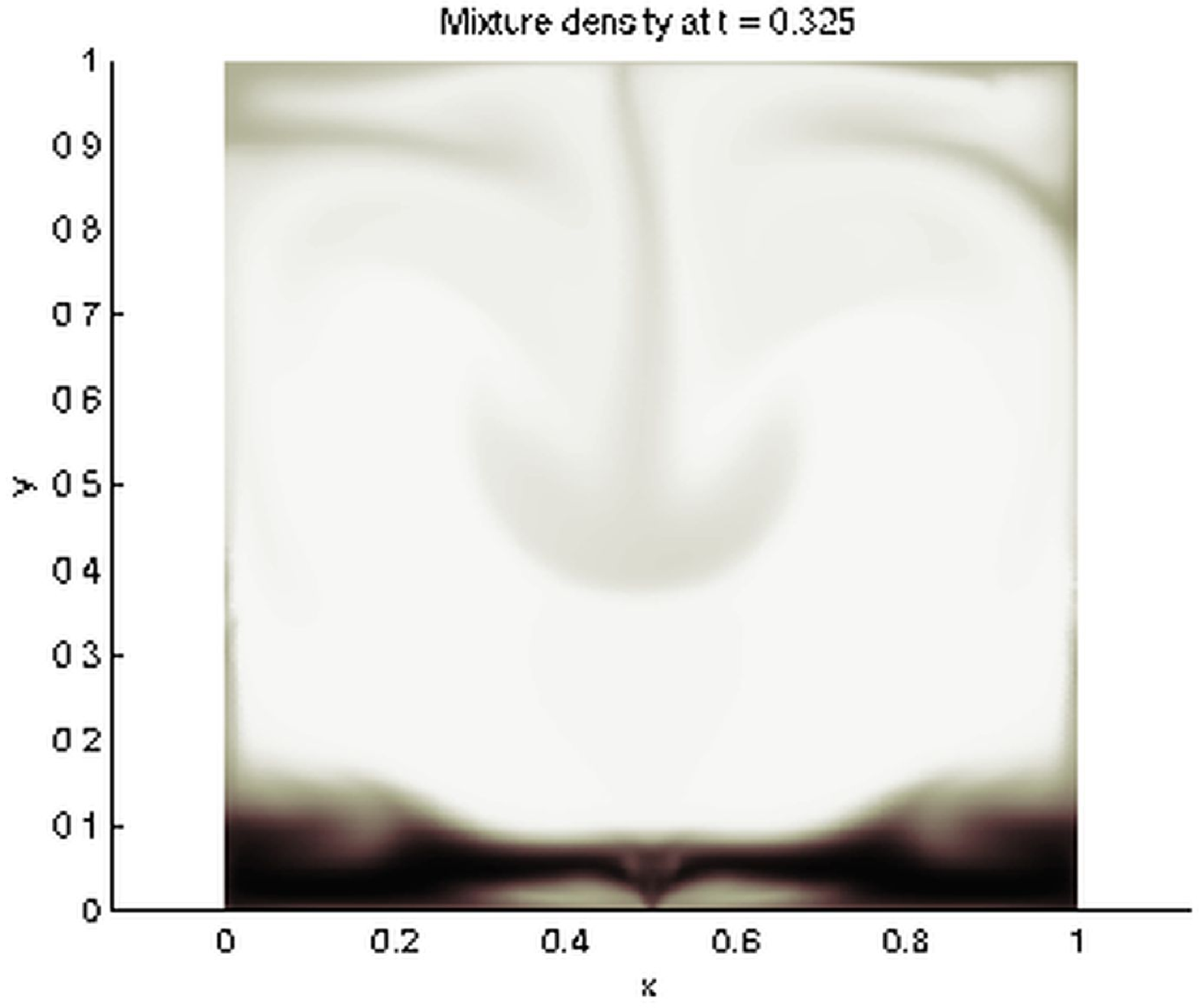}} \quad
	\subfigure[$t = 0.35$]%
	{\includegraphics[width=0.46\textwidth]{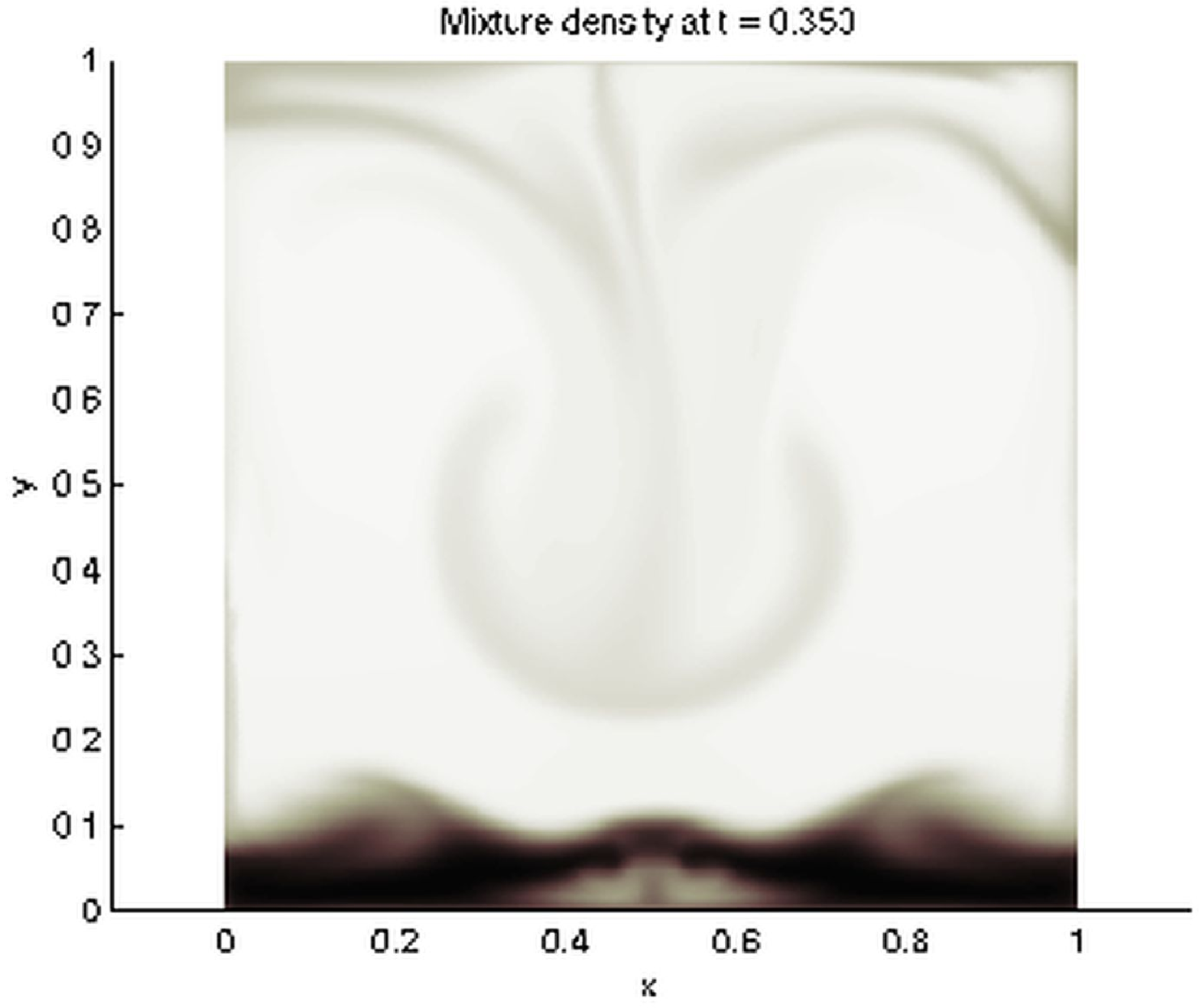}}
	\caption[Water drop test case. Side jets interflow at the center]{Water drop test case. Side jets flowing down the centerline.}
	\label{fig:DropAsym}
\end{figure}

\begin{figure}
	\centering
	\subfigure[$t = 0.4$]%
	{\includegraphics[width=0.46\textwidth]{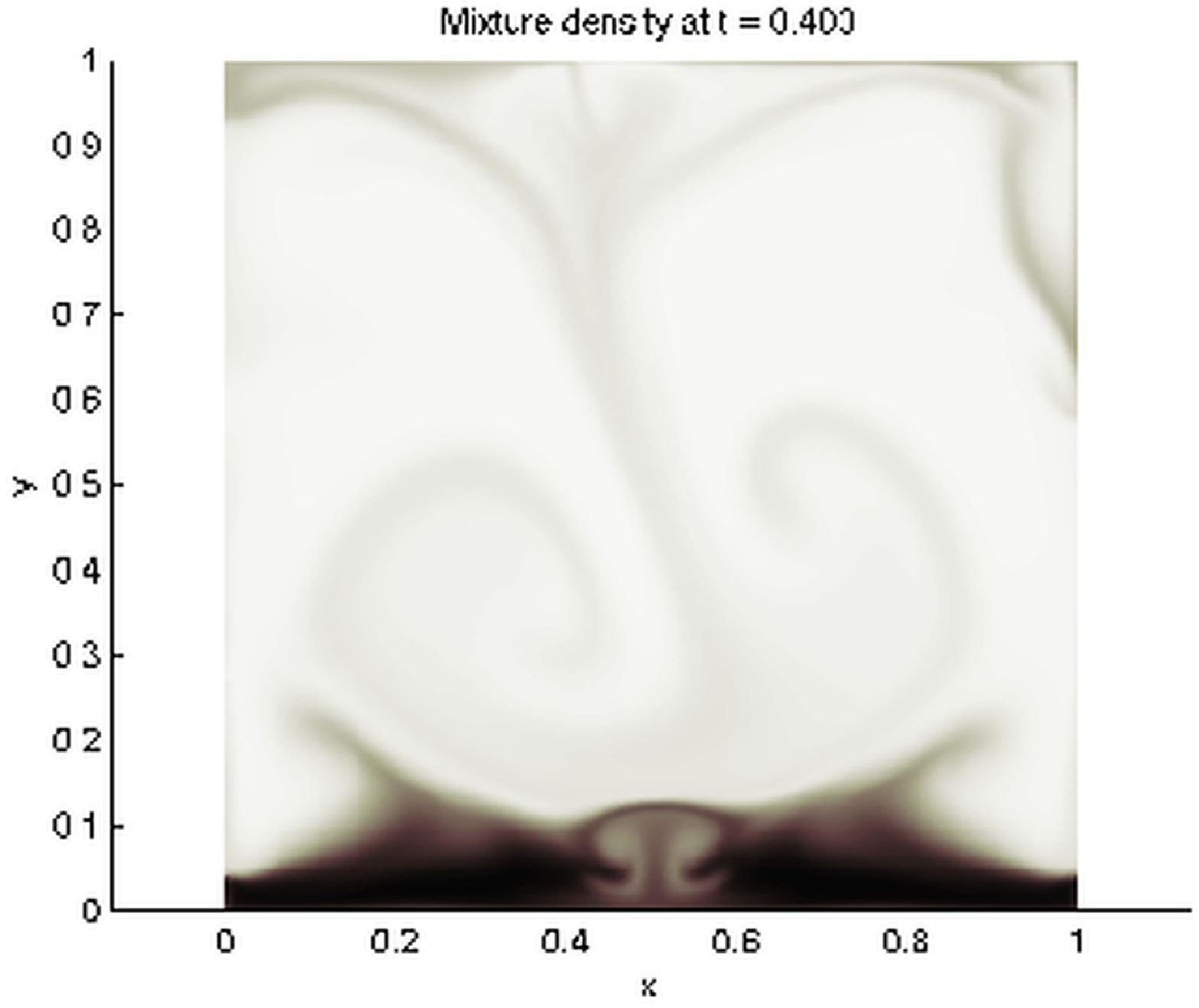}} \quad
	\subfigure[$t = 0.45$]%
	{\includegraphics[width=0.46\textwidth]{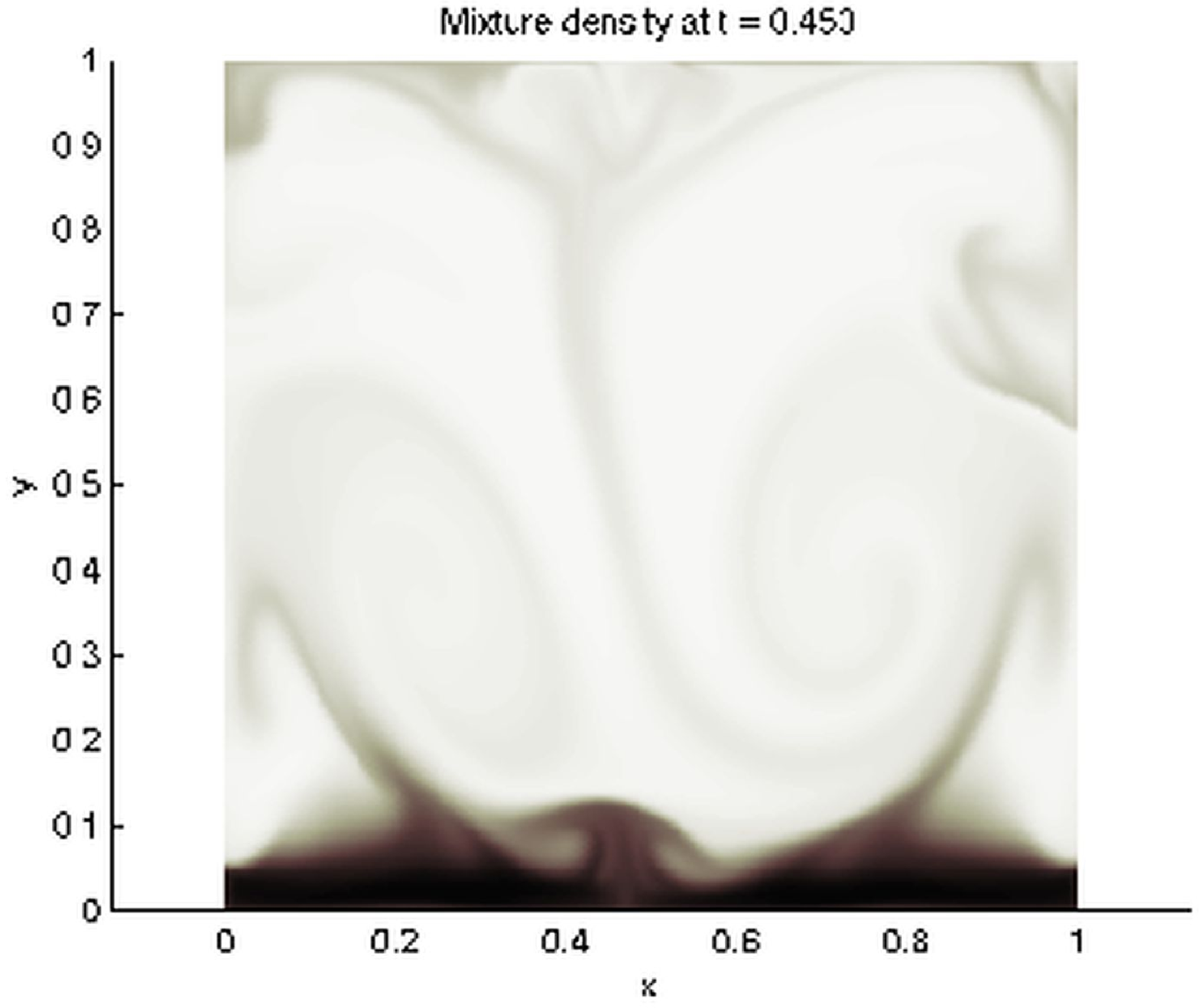}}
	\caption[Central jet reflection from the bottom]{Water drop test case. Central jet reflection from the bottom.}
	\label{fig:lastDrop}
\end{figure}

\begin{figure}
	\centering
		\includegraphics[width=0.7\textwidth]{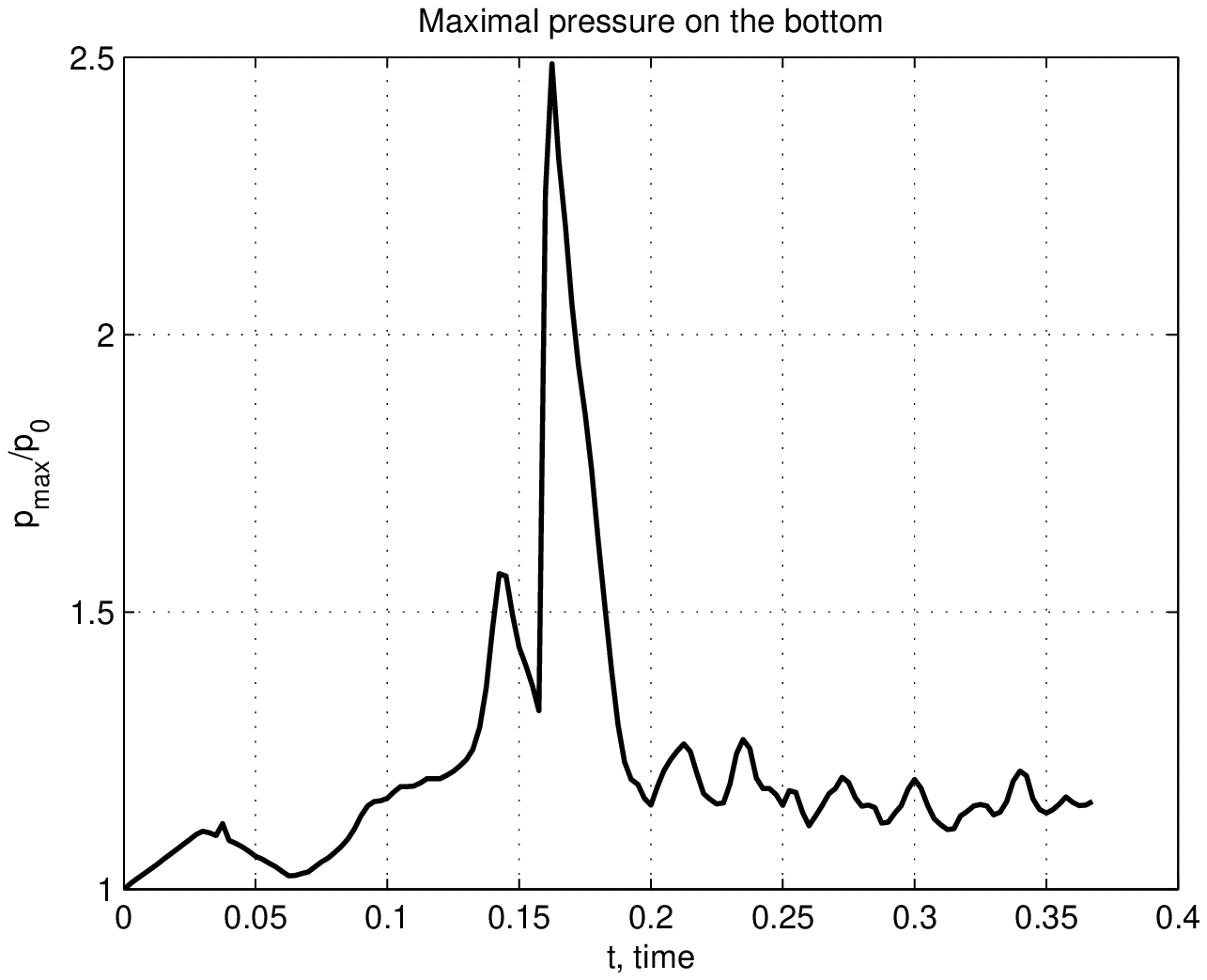}
	\caption{Water drop test case. Maximum bottom pressure as a function of time.}
	\label{fig:bottompress}
\end{figure}

The geometry and initial condition for this test case are shown on \figurename~\ref{fig:drop_water}. Initially the velocity field is taken to be zero. The values of the other parameters are given in Table \ref{tab:diphaseparams}. The mesh used in this computation contained about $92 000$ control volumes (again they were triangles). The results of this simulation are presented in Figures \ref{fig:debutDrop}--\ref{fig:lastDrop}. In \figurename~\ref{fig:bottompress} we plot the maximal pressure on the bottom as a function of time:
\begin{equation*}
  t \longmapsto \max_{(x,y) \in [0,1]\times 0} p(x,y,t).
\end{equation*}
It is clear that the pressure exerted on the bottom reaches $2.5p_0$ due to the drop impact at $t\approx 0.16$.

\begin{rem}
Beginning with \figurename~\ref{fig:DropAsym} one can see some asymmetry in the solution. It is not expected since the initial condition, computational domain and forcing term are fully symmetric with respect to the line $x=0.5$. This discrepancy is explained by the use of unstructured meshes in the computation. The arbitrariness of the orientation of the triangles introduces small perturbations which are sufficient to break the symmetry at the discrete level.
\end{rem}

\begin{figure}
	\centering
	\subfigure[$t = 0.1$]%
	{\includegraphics[width=0.46\textwidth]{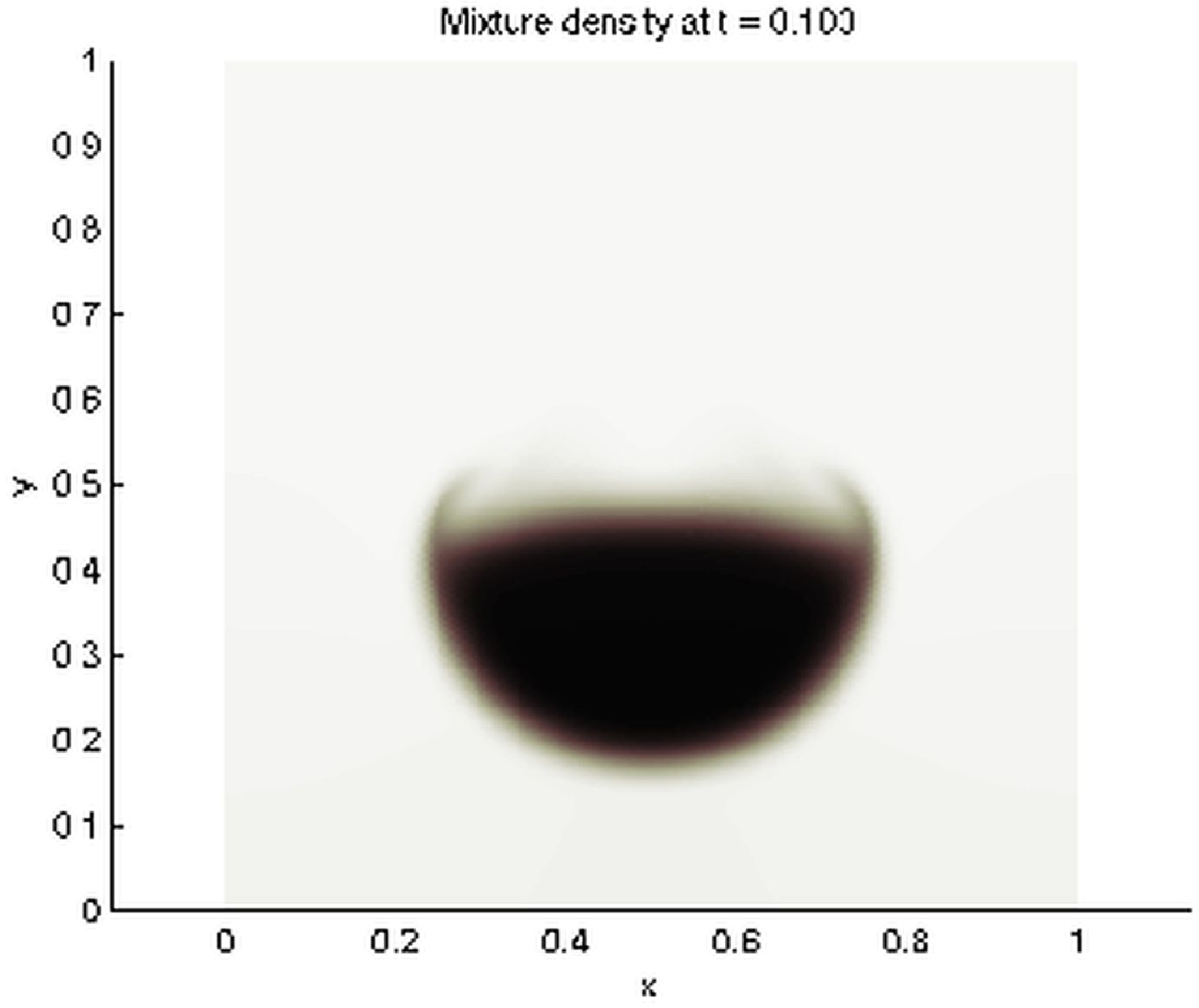}} \quad
	\subfigure[$t = 0.125$]%
	{\includegraphics[width=0.46\textwidth]{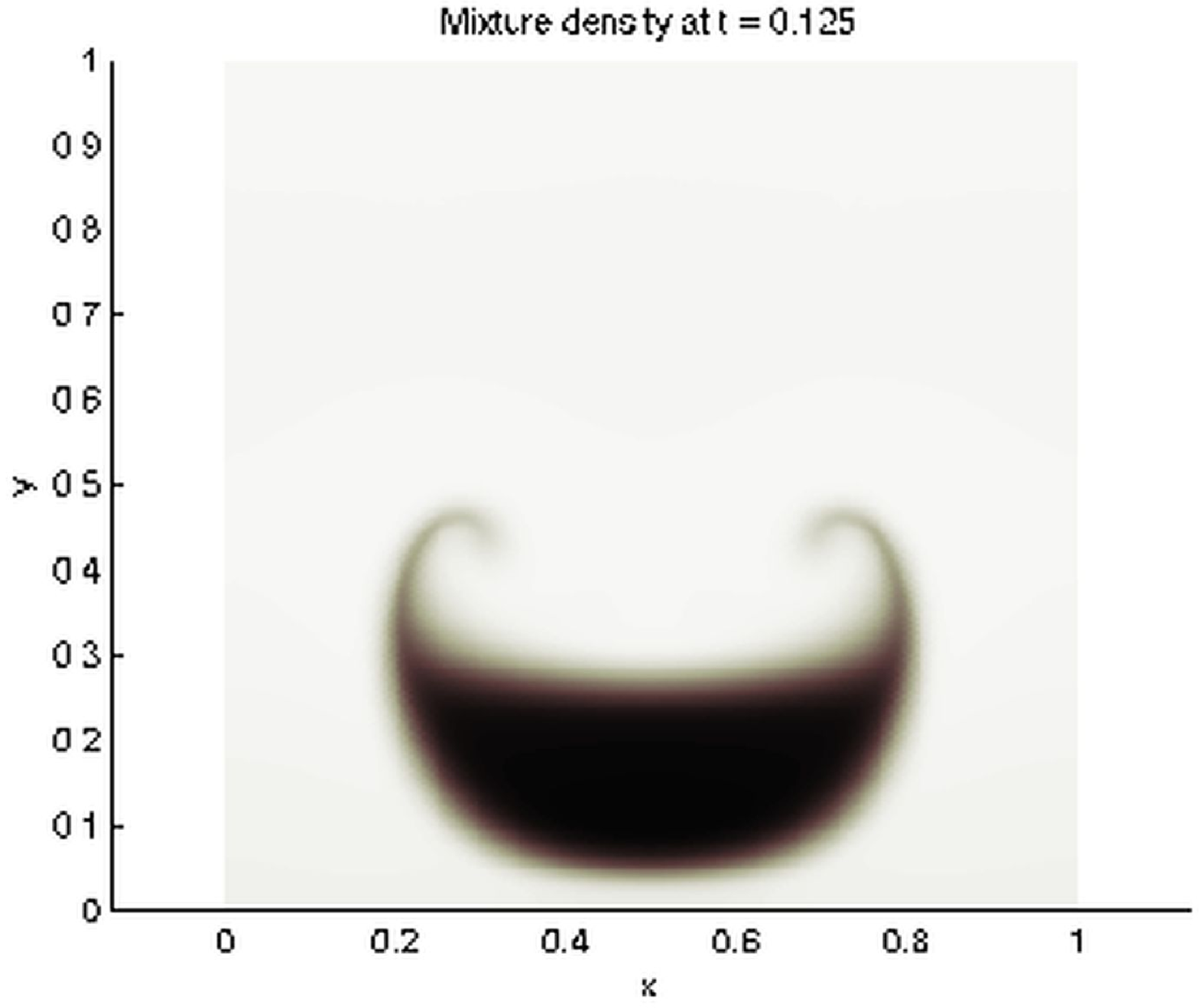}}
	\caption[Drop approaching container bottom]{Water drop test case. Drop approaching the bottom of the container.}
\end{figure}

\begin{figure}
	\centering
	\subfigure[$t = 0.135$]%
	{\includegraphics[width=0.46\textwidth]{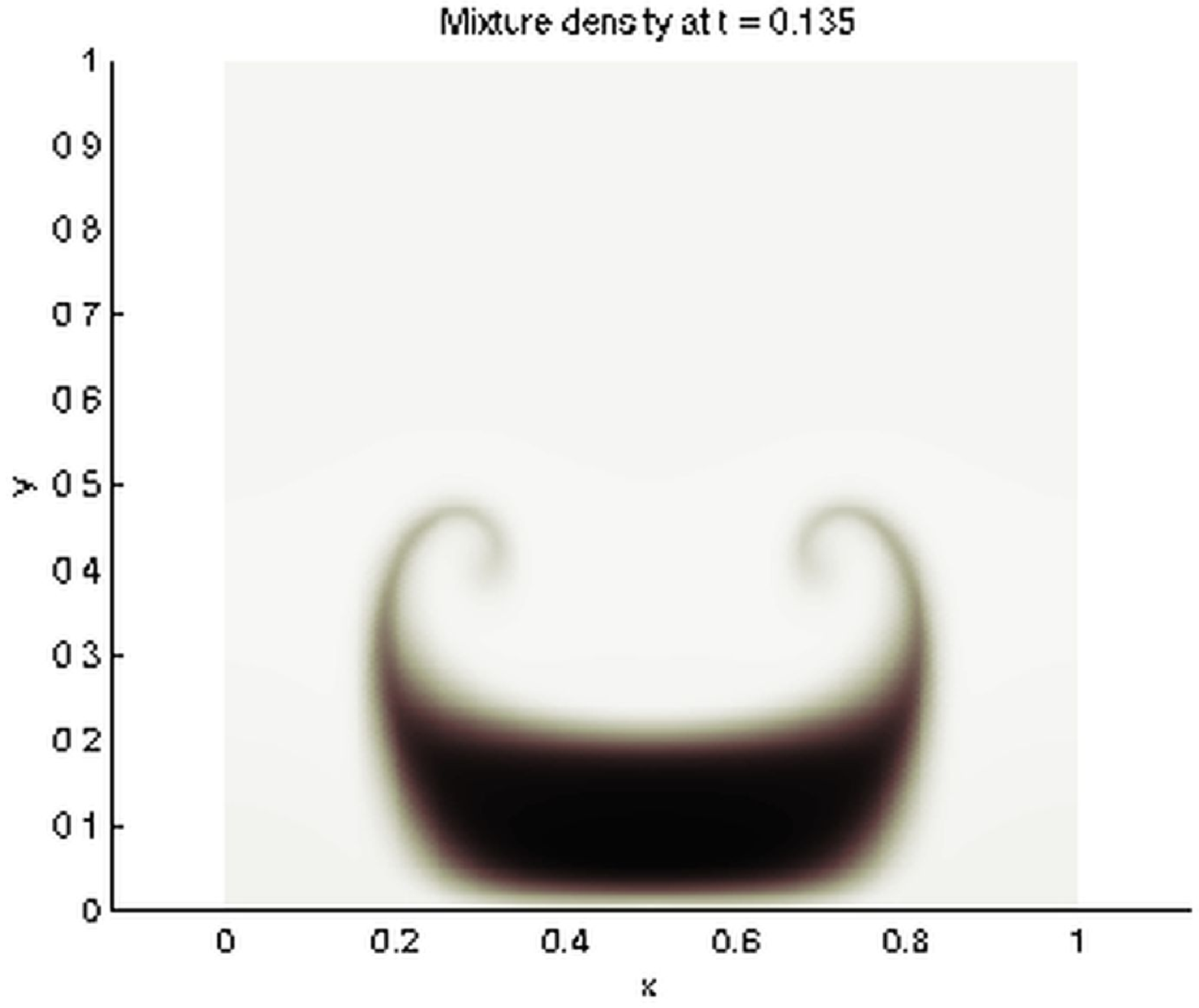}} \quad
	\subfigure[$t = 0.15$]%
	{\includegraphics[width=0.46\textwidth]{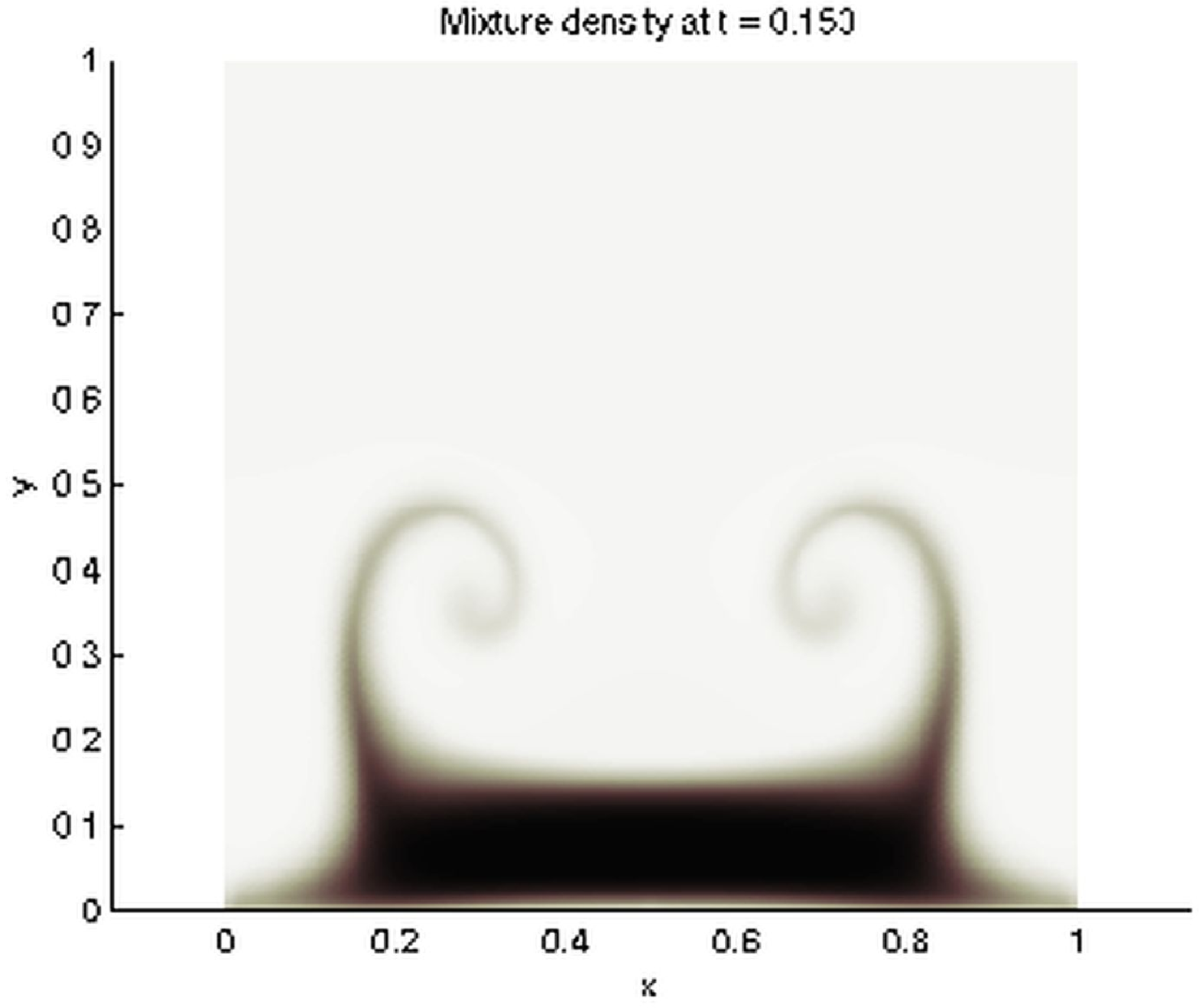}}
	\caption[Drop/bottom compressible interaction]{Water drop test case. Drop/bottom compressible interaction.}
\end{figure}

\begin{figure}
	\centering
	\subfigure[$t = 0.175$]%
	{\includegraphics[width=0.46\textwidth]{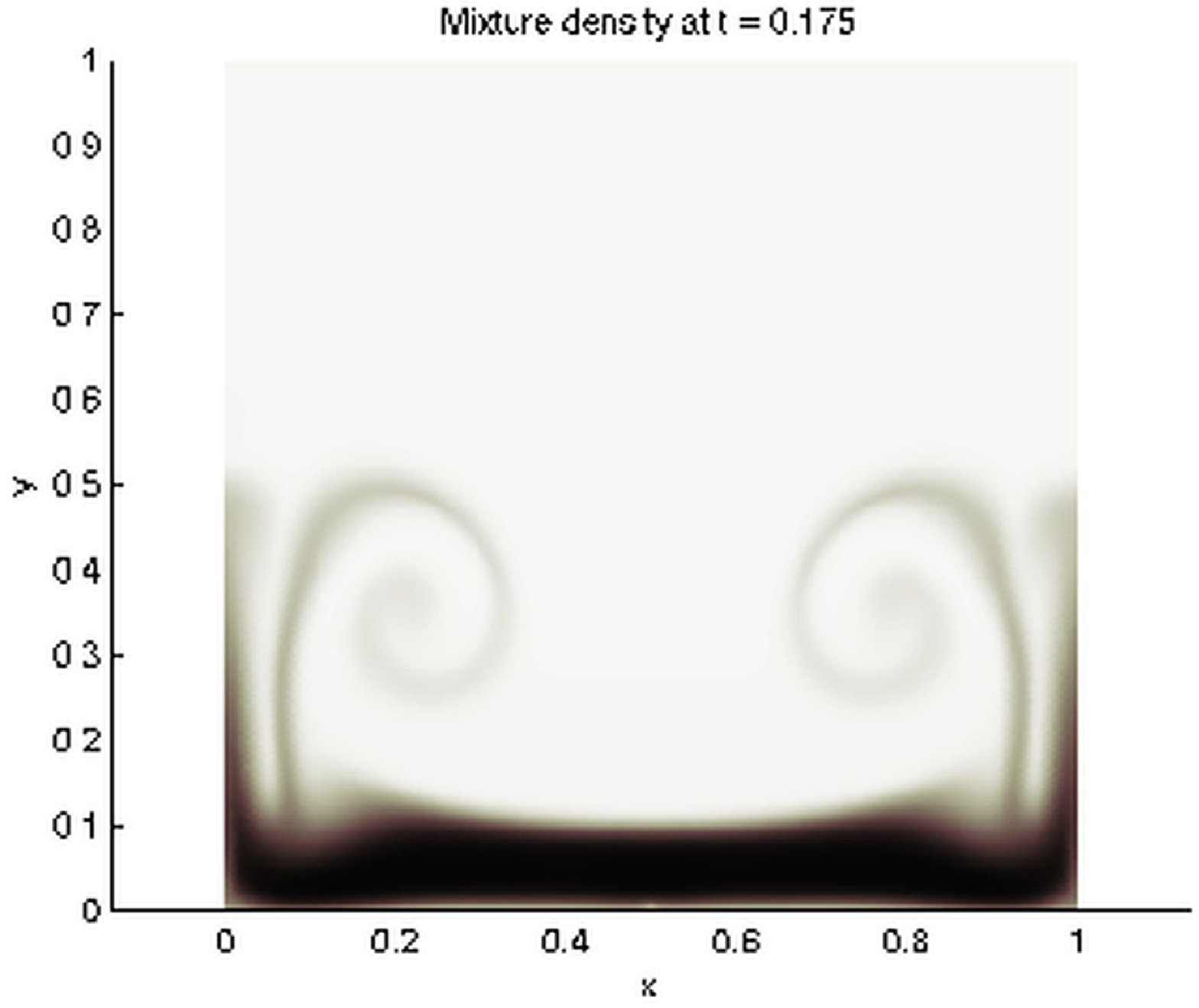}} \quad
	\subfigure[$t = 0.2$]%
	{\includegraphics[width=0.46\textwidth]{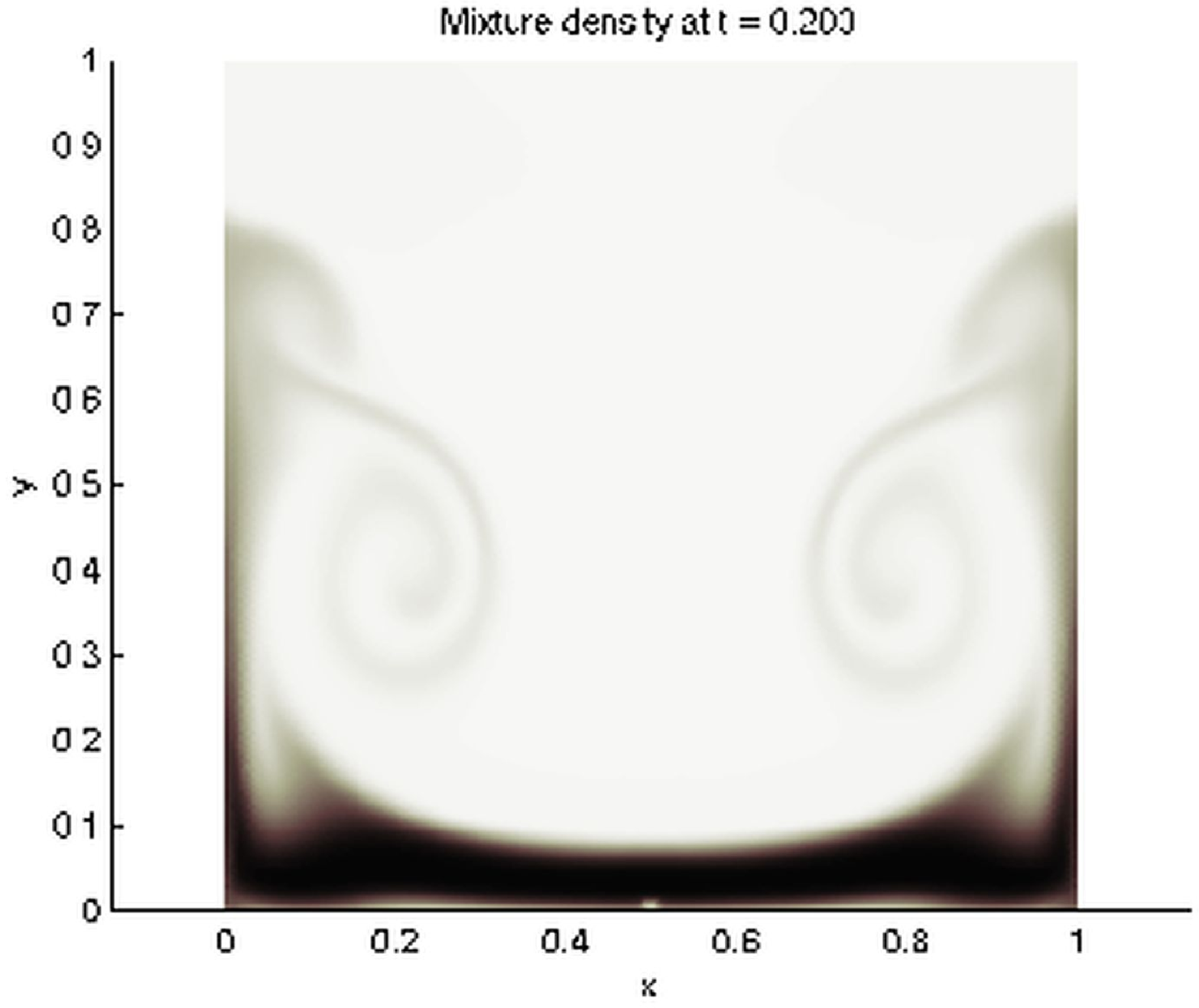}}
	\caption[Water drop test case. Vertical jets formation]{Water drop test case. Vertical jets formation.}
\end{figure}

\begin{figure}
	\centering
	\subfigure[$t = 0.225$]%
	{\includegraphics[width=0.46\textwidth]{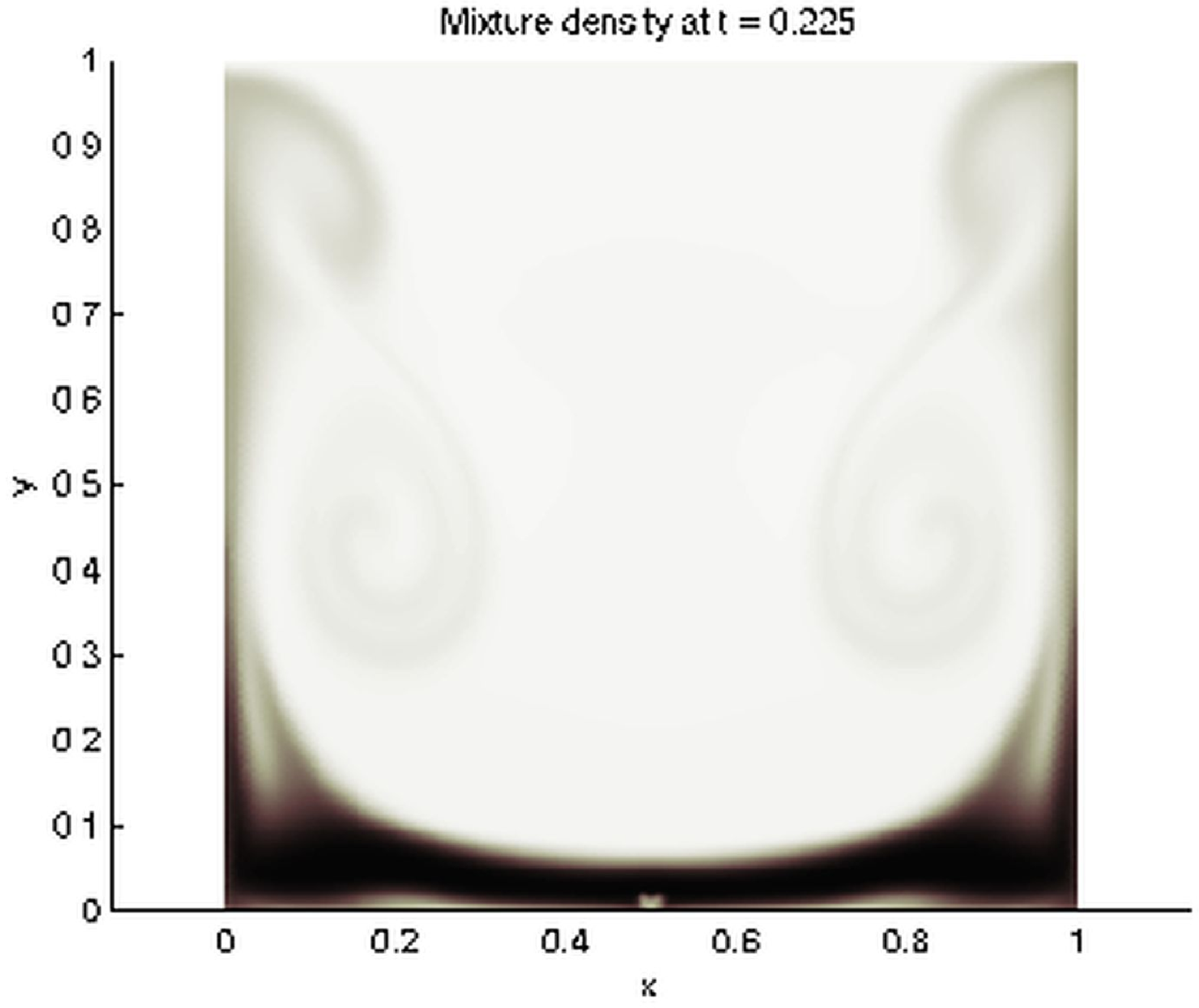}} \quad
	\subfigure[$t = 0.275$]%
	{\includegraphics[width=0.46\textwidth]{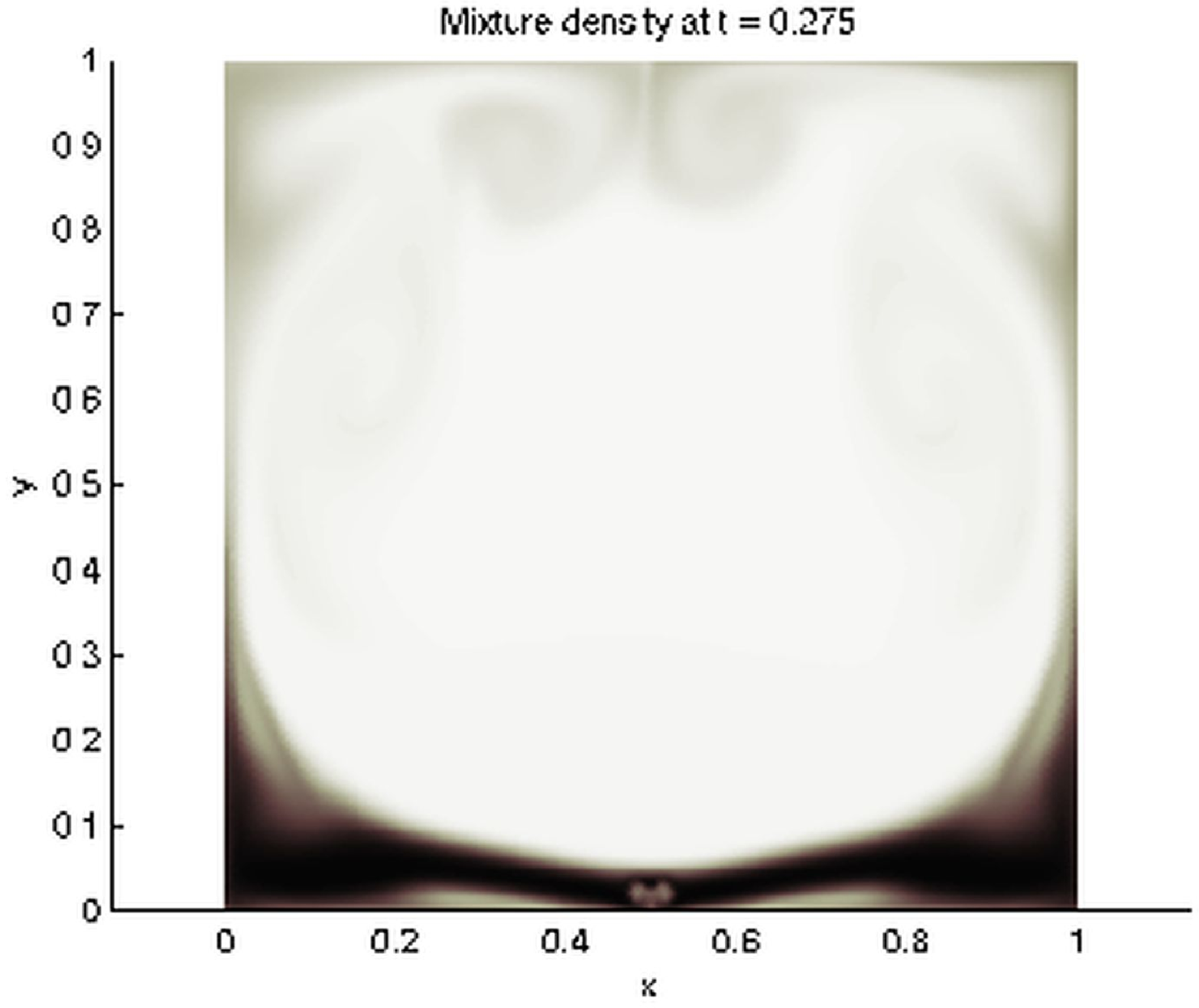}}
	\caption[Water drop test case. Side jets crossing]{Water drop test case. Side jets crossing.}
\end{figure}

\section{Conclusions}

In this article we have presented a simple mathematical model for simulating water wave impacts. Associated to this model, which avoids the costly capture of free surfaces, we have built a numerical solver which is: (i)~ second-order accurate on smooth solutions, (ii)~stable even for solutions with very strong gradients (and solutions with shocks) and (iii)~locally {\it exactly} conservative with respect to the mass of each fluid, momentum and total energy. This last property, (iii), which is certainly the most desirable from the physical point of view, is an immediate byproduct of our cell-centered finite volume method.

We have shown here the good behavior of this framework on simple test cases and we are presently working on quantitative comparisons in the context of real applications.

\appendix

\section{Some technical results}

  The constants $C_V^\pm$ can be calculated after simple algebraic manipulations of equations (\ref{eq:light}), (\ref{eq:heavy}) and matching with experimental values at normal conditions:
  \begin{equation*}
    C_V^- \equiv \frac{p_0}{(\gamma^- -1)\rho^-_0 T_0},
  \end{equation*}
  \begin{equation*}
    C_V^+ \equiv \frac{\gamma^+ p_0 + \pi^+}{(\gamma^+-1)\gamma^+\rho_0^+ T_0}.
  \end{equation*}
  For example, for an air/water mixture under normal conditions we have the values given in Table~\ref{tab:diphaseparams}.

\begin{table}
	\begin{center}
		\begin{tabular}{cc}
  		\hline\hline
  		\textit{parameter} & \textit{value} \\
      \hline
      	$p_0$ & $10^5$ $Pa$ \\
      \hline
        $\rho^+_0$ & $10^3$ $kg/m^3$ \\
      \hline
        $\rho^-_0$ & $1.29$ $kg/m^3$ \\
      \hline
        $T_0$ & $300$ $K$ \\
      \hline
        $\gamma^-$ & $1.4$ \\
      \hline
        $\gamma^+$ & $7$ \\
      \hline
        $\pi^+$ & $2.1\times 10^9$ $Pa$ \\
      \hline
        $C_V^+$ & $166.72$ $\frac{J}{kg\cdot K}$ \\
      \hline
        $C_V^-$ & $646.0$ $\frac{J}{kg\cdot K}$ \\
      \hline
        $g$ & $100$ $\frac{m}{s^2}$ \\
      \hline\hline
		\end{tabular}
	  \caption[Parameters for an air/water mixture under normal conditions]%
	  {Values of the parameters for an air/water mixture under normal conditions. The rather high value of the acceleration due to gravity does not correspond to any physical situation. Nevertheless, this value was chosen in order to accelerate the dynamic processes in the test cases.}
	  \label{tab:diphaseparams}
  \end{center}
\end{table}

The sound velocities in each phase are given by the following formulas:
\begin{equation}\label{eq:soundspeed}
  (c_s^-)^2 = \frac{\gamma^- p^-}{\rho^-}, \qquad
  (c_s^+)^2 = \frac{\gamma^+ p^+ + \pi^+}{\rho^+}.
\end{equation}

\section{Dispersion relation in the pure fluid limit}
\label{pure}

Let us provide the dispersion relation for waves propagating along the interface in the limit of two superposed 
compressible heavy fluids.
Consider the linearization of the equations around the equilibrium state
\begin{eqnarray*}
\underline{\eta}(\x,t) & = & 0, \\
\underline{\rho}^\pm(\x,z,t) & = & \rho_0^\pm \exp\left(-\frac{gz}{(c_s^\pm)^2}\right), \\
\underline{\u}^\pm(\x,z,t) & = & \vec{0}, \\
\underline{p}^\pm(\x, z, t) & = & (c_s^\pm)^2 \rho_0^\pm \exp\left(-\frac{gz}{(c_s^\pm)^2}\right) - (c_s^\pm)^2 \rho_0^\pm + p_0, \\
\underline{s}^\pm & = & s_0^\pm
\label{equations-cc}
\end{eqnarray*}
If we assume that ${gz}/{(c_s^\pm)^2}$ is small, an approximation to the equilibrium state is given by
\begin{eqnarray}
\underline{\eta}(\x,t) & = & 0, \\
\underline{\rho}^\pm(\x,z,t) & = & \rho_0^\pm, \\
\underline{\u}^\pm(\x,z,t) & = & \vec{0}, \\
\underline{p}^\pm(\x,z,t) & = & p_0 - \rho_0^\pm gz, \\
\underline{s}^\pm & = & s_0^\pm
\label{equations-ccs}
\end{eqnarray}

We write the following perturbations of the equilibrium state:
$$
\eta = \underline{\eta} + \zeta + \ldots, \quad
\rho^\pm = \underline{\rho}^\pm + \varrho^\pm + \ldots, \quad
\u^\pm = \vec{0} + \v^\pm + \ldots, \quad
p^\pm = \underline{p}^\pm + q^\pm + \ldots,
$$
with in addition $s^\pm = s_0^\pm + \sigma^\pm$.

The linearized equations for both fluids (\ref{linclass1})--(\ref{linclass3}) read
\begin{eqnarray}
\frac{\partial\varrho^\pm}{\partial t} + \rho_0^\pm \nabla_h\cdot\v_h^\pm
+ \rho_0^\pm \frac{\partial w^\pm}{\partial z} & = & 0, \\
\frac{\partial \v^{\pm}_h}{\partial t} + \frac{1}{\rho_0^\pm} \nabla_h q^\pm &=& 0, \\
\frac{\partial w^{\pm}}{\partial t} + \frac{1}{\rho_0^\pm} \frac{\partial q^\pm}{\partial z} &=& 0, \\
\frac{\partial q^\pm}{\partial t} &=& \left(c_s^\pm\right)^2 \frac{\partial \varrho^\pm}{\partial t}.
\end{eqnarray}

The kinematic and dynamic boundary conditions along the interface are
\begin{eqnarray}
\frac{\partial\zeta}{\partial t} &=& w^{\pm}(\x,0,t), \\
q^+(\x,0,t) - \rho_0^+ g\zeta & = & q^-(\x,0,t) - \rho_0^- g\zeta.
\end{eqnarray}

We perform a classical perturbation analysis: we look for solutions in the form
\begin{equation}
  \begin{pmatrix}
        \varrho^\pm \\ \v^{\pm}_h \\ w^{\pm} \\ q^\pm
  \end{pmatrix} =
  \begin{pmatrix}
    \hat \varrho^\pm \\
    \hat \v^{\pm}_h \\
    \hat w^{\pm} \\
    \hat q^\pm
  \end{pmatrix} (z) e^{i(\k\cdot\x - \omega t)},
\end{equation}
that is periodic perturbations with wave number $\k$ and angular frequency $\omega$. One eventually obtains the following second-order ODE for $\hat q^\pm(z)$:
\begin{equation} \label{ode_pressure}
  \od{^2 \hat q^\pm}{z^2} - \left(|\k|^2 - \frac{\omega^2}{(c_s^\pm)^2}\right)\hat q^\pm = 0.
\end{equation}

Having $q^\pm$, one can find $\v^\pm$, $\varrho^\pm$ and $\zeta$. Assume a geometry of total depth $D$ bounded above and below by rigid walls located at $z=\alpha_0^-D$ and $z=-\alpha_0^+D$ respectively. The boundary conditions along the horizontal walls are
\begin{equation}
	w^\pm(\x,\mp\alpha_0^\pm D,t) = 0.
\end{equation}
Satisfying the kinematic and dynamic boundary conditions along the interface provides a solvability condition. In other words, the solution to the linearized problem provides the dispersion relation $\omega(\k)$, which shows the presence of two kinds of modes: the gravity modes and the acoustic modes.

The ODE for $\hat q^\pm$ shows that the sign of $\omega^2-|\k|^2(c_s^\pm)^2$ plays an
important role. There are three cases: both signs are negative, one is positive and one is negative, both signs are positive.

\subsection{Case where $c_s^-|\k| < \omega < c_s^+ |\k|$}
The assumption is based on the fact that the speed of sound $c_s^+$ is much higher than the
speed of sound $c_s^-$.
The solution of the linearized problem is
\begin{eqnarray*}
q^- & = & A \cos\left[ T_\omega^- |\k|(z-\alpha_0^- D) \right]{\rm e}^{i(\k\cdot\x-\omega t)}, \\
w^- & = & i A \frac{|\k|}{\omega\rho_0^-} T_\omega^- \sin\left[ T_\omega^- |\k|(z-\alpha_0^- D) \right]{\rm e}^{i(\k\cdot\x-\omega t)}, \\
\v^- & = & A \frac{\k}{\omega\rho_0^-} \cos\left[ T_\omega^- |\k|(z-\alpha_0^- D) \right]{\rm e}^{i(\k\cdot\x-\omega t)}, \\
\varrho^- & = & A \frac{1}{\left(c_s^-\right)^2} \cos\left[ T_\omega^- |\k|(z-\alpha_0^- D) \right]{\rm e}^{i(\k\cdot\x-\omega t)}, \\
\zeta & = & A \frac{|\k|}{\omega^2\rho_0^-} T_\omega^- \sin(T_\omega^- |\k|\alpha_0^- D){\rm e}^{i(\k\cdot\x-\omega t)},
\end{eqnarray*}
and
\begin{eqnarray*}
q^+ & = & B \cosh\left[ S_\omega^+ |\k|(z+\alpha_0^+ D) \right]{\rm e}^{i(\k\cdot\x-\omega t)}, \\
w^+ & = & -i B \frac{|\k|}{\omega\rho_0^+} S_\omega^+ \sinh\left[ S_\omega^+ |\k|(z+\alpha_0^+ D) \right]{\rm e}^{i(\k\cdot\x-\omega t)}, \\
\v^+ & = & B \frac{\k}{\omega\rho_0^+} \cosh\left[ S_\omega^+ |\k|(z+\alpha_0^+ D) \right]{\rm e}^{i(\k\cdot\x-\omega t)}, \\
\varrho^+ & = & B \frac{1}{\left(c_s^+\right)^2} \cosh\left[ S_\omega^+ |\k|(z+\alpha_0^+ D) \right]{\rm e}^{i(\k\cdot\x-\omega t)}, \\
\zeta & = & B \frac{|\k|}{\omega^2\rho_0^+} S_\omega^+ \sinh(S_\omega^+ |\k|\alpha_0^+ D){\rm e}^{i(\k\cdot\x-\omega t)}.
\end{eqnarray*}

Writing that both expressions for $\zeta$ are equal and satisfying the dynamic condition along the interface yields
a system of two homogeneous linear equations in $A$ and $B$. The dispersion relation for the acoustic modes is obtained
by setting the determinant equal to 0:
\begin{multline}
	\frac{\omega^2}{g|\k|}\left(
	T_\omega^-\tan(T_\omega^-\alpha_0^- |\k|D) - \theta S_\omega^+\tanh(S_\omega^+\alpha_0^+ |\k|D)
	\right) =\\=
	(1-\theta)S_\omega^+T_\omega^-\tan(T_\omega^-\alpha_0^- |\k|D)\tanh(S_\omega^+\alpha_0^+ |\k|D),
\label{acou}
\end{multline}
where
$ \theta := {\rho_0^-}/{\rho_0^+}.$

\subsection{Case where $\omega < \min(c_s^-|\k|,c_s^+ |\k|)$}

Assume now that $\omega < c_s^-|\k|$.
The solution of the linearized problem is unchanged for the liquid. For the gas it becomes
\begin{eqnarray*}
q^- & = & A \cosh\left[ S_\omega^- |\k|(z-\alpha_0^- D) \right], \\
w^- & = & -i A \frac{|\k|}{\omega\rho_0^-} S_\omega \sinh\left[ S_\omega^- |\k|(z-\alpha_0^- D) \right], \\
\v^- & = & A \frac{\k}{\omega\rho_0^-} \cosh\left[ S_\omega^- |\k|(z-\alpha_0^- D) \right], \\
\varrho^- & = & A \frac{1}{c_0^2} \cosh\left[ S_\omega^- |\k|(z-\alpha_0^- D) \right], \\
\zeta & = & - A \frac{|\k|}{\omega^2\rho_0^-} S_\omega^- \sinh(S_\omega^- |\k|\alpha_0^- D).
\end{eqnarray*}

Writing that both expressions for $\zeta$ are equal and satisfying the dynamic condition along the interface yields
a system of two homogeneous linear equations in $A$ and $B$. The dispersion relation for the gravity modes is obtained
by setting the determinant equal to 0:
\begin{multline}
  \frac{\omega^2}{g|\k|}\left(
  S_\omega^-\tanh(S_\omega^-\alpha_0^- |\k|D) +
  \theta S_\omega^+\tanh(S_\omega^+\alpha_0^+ |\k|D)\right) = \\ =
  (1-\theta)S_\omega^- S_\omega^+\tanh(S_\omega^-\alpha_0^-|\k|D) \tanh(S_\omega^+\alpha_0^+|\k|D).
\label{gravity}
\end{multline}

\subsection{Case where $\omega > \max(c_s^+|\k|,c_s^- |\k|)$}

Assume now that $\omega > c_s^+|\k|$.
The solution of the linearized problem for the liquid becomes
\begin{eqnarray*}
q^+ & = & B \cos\left[ T_\omega^+ |\k|(z+\alpha_0^+ D) \right]{\rm e}^{i(\k\cdot\x-\omega t)}, \\
w^+ & = & i B \frac{|\k|}{\omega\rho_0^+} T_\omega^+ \sin\left[ T_\omega^+ |\k|(z+\alpha_0^+ D) \right]{\rm e}^{i(\k\cdot\x-\omega t)}, \\
\v^+ & = & B \frac{\k}{\omega\rho_0^+} \cos\left[ T_\omega^+ |\k|(z+\alpha_0^+ D) \right]{\rm e}^{i(\k\cdot\x-\omega t)}, \\
\varrho^+ & = & B \frac{1}{\left(c_s^+\right)^2} \cos\left[ T_\omega^+ |\k|(z+\alpha_0^+ D) \right]{\rm e}^{i(\k\cdot\x-\omega t)}, \\
\zeta & = & -B \frac{|\k|}{\omega^2\rho_0^+} T_\omega^+ \sin(T_\omega^+ |\k|\alpha_0^+ D){\rm e}^{i(\k\cdot\x-\omega t)}.
\end{eqnarray*}
Writing that both expressions for $\zeta$ are equal and satisfying the dynamic condition along the interface yields
a system of two homogeneous linear equations in $A$ and $B$. The dispersion relation for the acoustic modes is obtained
by setting the determinant equal to 0:
\begin{multline}
	\frac{\omega^2}{g|\k|}\left(
	T_\omega^-\tan(T_\omega^-\alpha_0^- |\k|D) + \theta T_\omega^+\tan(T_\omega^+\alpha_0^+ |\k|D)
	\right) =\\=
	-(1-\theta)T_\omega^+T_\omega^-\tan(T_\omega^-\alpha_0^- |\k|D)\tan(T_\omega^+\alpha_0^+ |\k|D).
\label{acoubis}
\end{multline}

\section{Isentropic flows}

Let us consider the isentropic version of the system of equations
(\ref{eq:massphys})--(\ref{eq:energyphys}). It reads
\begin{eqnarray}\label{eq:massphys_ise}
(\alpha^+\rho^+)_t  + \div(\alpha^+\rho^+\u) &=& 0, \\
\label{eq:massphys2_ise}
  (\alpha^-\rho^-)_t  + \div(\alpha^-\rho^-\u) &=& 0, \\
  (\rho\u)_t + \div\left(\rho\u\otimes\u + p\I\right) &=& \rho\g,
\end{eqnarray}
with the equation of state
\begin{equation}\label{EOSise}
p=\mathcal{P}_I(\alpha, \rho)\,,
\end{equation}
where the subscript $I$ stands for isentropic.

One can determine $\mathcal{P}_I$ as follows. First consider the two equations
\begin{eqnarray}
  (1+\alpha)\rho^++(1-\alpha)\rho^- &=& 2\rho\,, \\
  \mathcal{P}_I^+(\rho^+)-\mathcal{P}_I^-(\rho^-) &=& 0\,.
\end{eqnarray}
Given $p>0$, we denote by $\mathcal{R}_I^\pm(p)$ the solutions $\rho^\pm$ to
\begin{equation}
\mathcal{P}_I^\pm(\rho^\pm)=p\,.
\end{equation}
It follows that
\begin{equation}
\rho = \frac{1+\alpha}{2}\mathcal{R}_I^+(p)+\frac{1-\alpha}{2}\mathcal{R}_I^-(p)\,.
\end{equation}
The inversion of this equation leads to $p=\mathcal{P}_I(\alpha,\rho)$, which is the equation of state that appears
in (\ref{EOSise}).

In order to see if one can go further analytically, let us consider the particular case of stiffened gases. The equations of state are
\begin{equation}\label{EOS567ise}
p^\pm + \pi^\pm = (\gamma^\pm-1) \rho^\pm e^\pm\,,
\end{equation}
and
$$ e^\pm = C_V^\pm T^\pm + \frac{\pi^\pm}{\gamma^\pm \rho^\pm}. $$

\begin{proposition}
Possible entropies $s^\pm$ for stiffened gases are given by
\begin{equation}
s^\pm = C_V^\pm \log \left( \frac{p^\pm + \pi^\pm / \gamma^\pm}{(\rho^\pm)^{\gamma^\pm}} \right)\,.
\end{equation}
\end{proposition}
This expression can be easily obtained by integrating the well-known differential relation
$$ T \mbox{d}s = \mbox{d}e + p\,\mbox{d}\left(\frac{1}{\rho}\right). $$

\begin{remark}
Other possible entropies $s^\pm$ for stiffened gases, which differ from the previous ones by a constant, are given by
\begin{equation}
s^\pm = C_V^\pm \log \left( \frac{e^\pm - \pi^\pm / \gamma^\pm \rho^\pm}{(\rho^\pm)^{\gamma^\pm-1}} \right)\,.
\end{equation}
\end{remark}
Thus, saying that $s^\pm$ is constant boils down to saying that

\begin{equation} \label{eise}
e^\pm = \frac{\pi^\pm}{\gamma^\pm \rho^\pm} + \frac{A^\pm}{\gamma^\pm-1}(\rho^\pm)^{\gamma^\pm-1}\,,
\end{equation}
where $A^\pm$ is a constant. Substituting (\ref{eise}) into the EOS (\ref{EOS567ise}) yields
\begin{equation}\label{EOSise2}
p^\pm + \frac{\pi^\pm}{\gamma^\pm} =  A^\pm (\rho^\pm)^{\gamma^\pm}\,,
\end{equation}
that is
\begin{equation}\label{EOSise3}
\mathcal{P}_I^\pm(\rho^\pm) = - \frac{\pi^\pm}{\gamma^\pm} +  A^\pm (\rho^\pm)^{\gamma^\pm}\,.
\end{equation}
Thus
\begin{equation}\label{rhoise}
\mathcal{R}_I^\pm(p) = \left(\frac{p+{\pi^\pm}/{\gamma^\pm}}{A^\pm}\right)^{1/{\gamma^\pm}}\,,
\end{equation}
and the equation which gives $\mathcal{P}_I(\alpha,\rho)$ is
\begin{equation}\label{rhoise2}
\rho = \frac{1+\alpha}{2}\left(\frac{p+{\pi^+}/{\gamma^+}}{A^+}\right)^{1/{\gamma^+}}
+\frac{1-\alpha}{2}\left(\frac{p+{\pi^-}/{\gamma^-}}{A^-}\right)^{1/{\gamma^-}}\,.
\end{equation}
Even in the special case of two perfect gases where $\pi^\pm=0$, this equation is in general transcendental.
This is to be contrasted with the general case (the case with a variable entropy), where $\mathcal{P}(\alpha,\rho,e)$
can be calculated explicitly by algebraic equations.

From now on, we denote the set of equations (\ref{eq:massphys_ise})--(\ref{EOSise}) by (E$_I$).
In order to study small perturbations around basic smooth and stationary solutions, it
is more convenient to use the set of isentropic equations (E$_I$) rewritten
in the physical variables $\alpha$, $\u$, $p$.

These equations are given in the following proposition.
\begin{proposition}
The equivalent system to (E$_I$) in variables $\alpha$, $\u$, $p$ is
\begin{eqnarray}\label{eq:vitesseise}
\u_t + \u\cdot\nabla\u + \frac{1}{\rho}\nabla p &=& \g\,, \\\label{eq:pressionise}
  p_t + \u\cdot\nabla p + \rho c_{Is}^2\div \u &=& 0\,, \\\label{eq:alphaise}
  \alpha_t + \u\cdot\nabla \alpha + (1-\alpha^2)\,\delta_I\,\div \u &=& 0\,,
  \end{eqnarray}
where $c_{Is}^2$ and $\delta_I$ are given by

\begin{equation} \label{cis}
c_{Is}^2 = \frac{\Gamma^+ \Gamma^-}{\Gamma_0} \frac{\overline{\rho}\rho}{\rho^+\rho^-} \frac{p}{\rho}\,, \quad \Gamma_0 = \frac{1+\alpha}{2}\Gamma^-
+ \frac{1-\alpha}{2} \Gamma^+\,,
\end{equation}
\begin{equation} \label{gammai}
\Gamma^\pm \equiv \frac{\rho^\pm}{p} \frac{\mbox{d}\mathcal{P}_I^\pm(\rho^\pm)}{\mbox{d}\rho^\pm}=\gamma^\pm+\frac{\pi^\pm}{p}\,,
\quad \overline{\rho} = \alpha^-\rho^+ + \alpha^+\rho^-\,,
\end{equation}
\begin{equation} \label{deltai}
\delta_I = \frac{\Gamma^+-\Gamma^-}{2\Gamma_0}\,.
\end{equation}
\end{proposition}
\begin{remark}
In the one-fluid case (take for example $\alpha^+=1, \alpha^-=0, \alpha=1$), one finds
\begin{equation}
c_s = c_{Is} = c_s^+ \,,
\end{equation}
while, in the two-fluid case, $c_s \neq c_{Is}$.
\end{remark}

The analysis for the dispersion relation is quite similar to the general case. 
The steady state is denoted by $\underline{\alpha}^\pm$, $\underline{\rho}^\pm$, $\underline{p}$ and $\underline{\u}$. We look for a special class of solutions which are motionless, uniform in the horizontal coordinates and continuously stratified
in the vertical direction:
\begin{equation*}
  \underline{\alpha}^\pm = \underline{\alpha}^\pm (z), \quad
  \underline{\rho}^\pm = \underline{\rho}^\pm (z), \quad
  \underline{p} = \underline{p} (z), \quad
  \underline{\u} = \vec{0}.
\end{equation*}
Again we take $\underline{\rho}$ to be constant. One must solve
\begin{equation}
  (1+\underline{\alpha}(z))\mathcal{R}_I^+(\underline{p}(z)) + (1-\underline{\alpha}(z))\mathcal{R}_I^-(\underline{p}(z))
  = 2\rho_0\,,
\end{equation}
in order to find $\underline{\alpha}(z)$ and $\underline{\rho}^\pm$. It is easy to see that
\begin{equation}
  \underline{\alpha}(z) = \frac{2\rho_0 - \mathcal{R}_I^+(\underline{p}(z)) - \mathcal{R}_I^-(\underline{p}(z))}
{\mathcal{R}_I^+(\underline{p}(z)) - \mathcal{R}_I^-(\underline{p}(z))}\,,
\end{equation}
with $\underline{p}(z)$ given by (\ref{hydrop2}).

The analysis is the same as before, except that
$\underline{c}_s^2$ and $\underline{\delta}$ are replaced by the values for the isentropic case.

\newpage
\tableofcontents
\newpage
\bibliographystyle{plain}
\bibliography{biblio}
\end{document}